\pdfoutput=1
\documentclass[applsci,article,accept,moreauthors,pdftex,10pt,a4paper]{Definitions/mdpi}
\firstpage{1}
\pubvolume{8}
\issuenum{11}
\articlenumber{2321}
\pubyear{2018}
\copyrightyear{2018}
\history{Received: 17 October 2018; Accepted: 17 November 2018; Published: 21 November 2018}
\setitemize{parsep=6pt,itemsep=0pt,leftmargin=*,labelsep=5.5mm}
\setenumerate{parsep=6pt,itemsep=0pt,leftmargin=*,labelsep=5.5mm}
\setlist[description]{itemsep=0mm}

\usepackage[labelformat=simple]{subcaption}

\DeclareCaptionLabelFormat{subcaptionlabel}{\normalfont(\textbf{#2}\normalfont)}
\Title{Identifying Real Estate Opportunities Using Machine~Learning}

\Author{Alejandro Baldominos $^{1,2,}$*\orcidA{}, Iv\'an Blanco $^{3}$, Antonio Jos\'e Moreno $^{2}$, Rub\'en Iturrarte $^{2}$, \'Oscar~Bern\'ardez $^{2}$ and Carlos Afonso $^{2}$}
\AuthorNames{Alejandro Baldominos, Iv\'an Blanco, Antonio Jos\'e Moreno, Rub\'en Iturrarte, \'Oscar Bernv\'ardez and Carlos Afonso}

\address{%
$^{1}$ \quad Computer Science Department, Universidad Carlos III de Madrid, 28911 Legan\'es, Spain\\   
$^{2}$ \quad Artificial Intelligence Group, Rentier Token, 28050 Madrid, Spain; antonio.moreno@refplatform.com (A.J.M.); ruben.iturrarte@refplatform.com (R.I.); oscar.bernardez@refplatform.com (\'O.B.); carlos.afonso@refplatform.com (C.A.)\\
$^{3}$ \quad Finance Department, Colegio Universitario de Estudios Financieros, 28040 Madrid, Spain; ivan.blanco@cunef.edu}
\corres{Correspondence: abaldomi@inf.uc3m.es; Tel.: +34-91-624-6258}

\abstract{The real estate market is exposed to many fluctuations in prices because of existing correlations with many variables, some of which cannot be controlled or might even be unknown. Housing prices can increase rapidly (or in some cases, also drop very fast), yet the numerous listings available online where houses are sold or rented are not likely to be updated that often. In some cases, individuals interested in selling a house (or apartment) might include it in some online listing, and~forget about updating the price. In other cases, some individuals might be interested in deliberately setting a price below the market price in order to sell the home faster, for various reasons. In this paper, we aim at developing a machine learning application that identifies opportunities in the real estate market in real time, i.e., houses that are listed with a price substantially below the market price. This program can be useful for investors interested in the housing market. We have focused in a use case considering real estate assets located in the Salamanca district in Madrid (Spain) and listed in the most relevant Spanish online site for home sales and rentals. The application is formally implemented as a regression problem that tries to estimate the market price of a house given features retrieved from public online listings. For building this application, we have performed a~feature engineering stage in order to discover relevant features that allows for attaining a high predictive performance. Several machine learning algorithms have been tested, including regression trees, $k$-nearest neighbors, support vector machines and neural networks, identifying advantages and handicaps of each of them.}


\begin{document}
%
%
\section{Introduction}

The real estate market is rapidly evolving. A recent report published by MSCI, Inc. (formerly Morgan Stanley Capital InternaTional) estimates the size of the professionally managed real estate investment market in \$8.5 trillion in 2017, increasing a total of \$1.1 trillion since the previous year \cite{Teuben18}. Of course, the real market size is expected to be much larger when counting assets which are not professionally managed or that are not object of investment.

When looked from a macroeconomic perspective, there are many aspects that significantly drive the behavior of this market, such as demographics, interest rates, government regulation and, for short, global economic health.

However, looking at the market evolution from a global perspective turns out to be too simplistic. Although the market at a global scale is very tightly correlated, there are many aspects influencing the behavior of markets at a local scale, such as political instability or the emergence of highly demanded ``hot spots'' that can shift rapidly. In addition, different market segments evolve at different paces, such as high-end luxury condos.

\begin{figure}[H]
\centering
\includegraphics[width=.8\columnwidth]{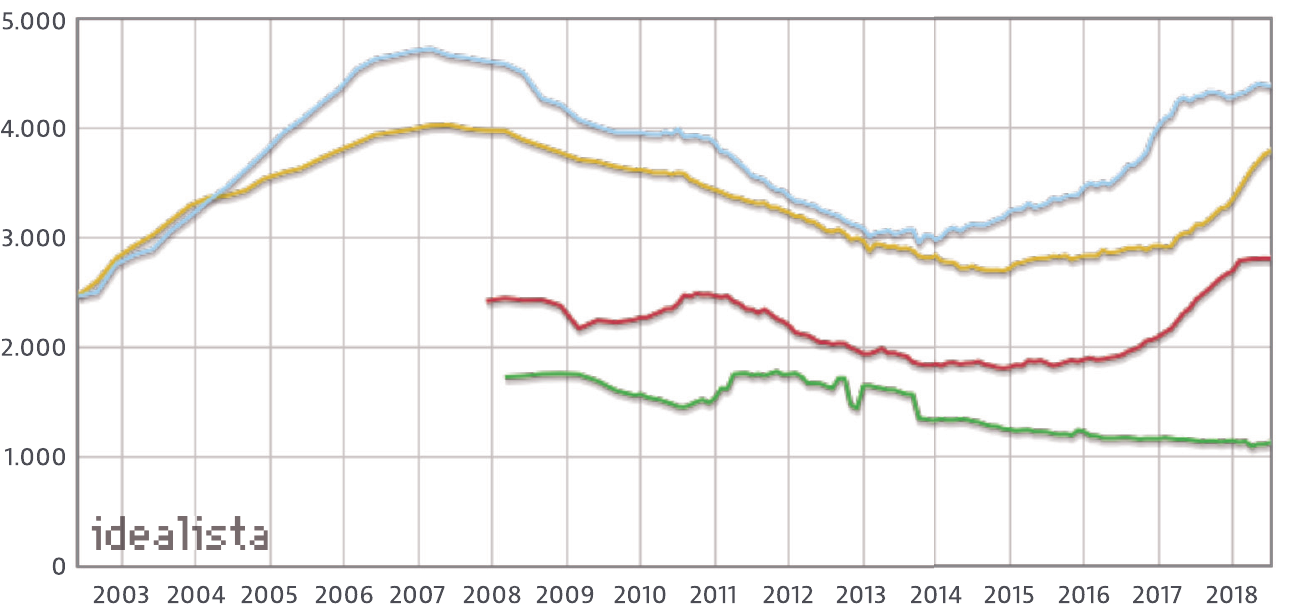}
\caption{Evolution of the Spanish resale real estate market, focusing on four different regions: Barcelona (blue), Madrid (yellow), Palma de Mallorca (red) and Lugo (green). Source: Idealista \cite{Idealista18}, reproduced with permission from Idealista.}
\label{fig:remarket_spain}
\end{figure}

An example of such differences can be observed in Figure \ref{fig:remarket_spain}, which shows the evolution of the price (measured in euros per square meter) of resale houses in four different Spanish regions: Barcelona (blue), Madrid (yellow), Palma de Mallorca (red) and Lugo (green). Data for Barcelona and Madrid are available since 2002, whereas, for Palma de Mallorca and Lugo, they are available after 2008.

Attending to the figure, we can see some common patterns in the evolution of Barcelona's and Madrid's markets, which are the two largest metropolitan areas in Spain: a growth in the early 2000s with a maximum in 2007, followed by a fall after 2008 due to the global financial crisis, which lasted until 2015, a moment after which prices have started to recover almost reaching maximum values again as of 2018.

Meanwhile, in Palma de Mallorca, which is a remarkable vacational place, the increase in the last two years is more pronounced than in Madrid and Barcelona. On the other hand, in Lugo, which is occupied mostly by rural areas, prices increased slightly in 2011, but are steadily decreasing since then. The figure clearly shows how although some global patterns can be found in the evolution, each region still shows some specificities in the prices evolution.

Of course, higher variability can be found when looking at specific assets. Most important factors driving the value of a house are the size and the location, but there are many other variables that are often taken into account when determining its value: number of bedrooms, proximity to some form of public transport (buses, underground, etc.), number and quality of schools in the area, shopping opportunities, availability of an elevator (in apartments located in higher floors), availability of gardens or parks, etc. However, certainly, the main force ultimately determining the value of houses is demand.

With such a high variability and unpredictable factors (e.g., one neighborhood deemed better or more fashionable than others), it is likely that the price of some assets will deviate from its expected value. When the actual price is much less than the expected value, we could be dealing with an~investment opportunity: an asset that could generate immediate profit if sold soon after its purchase.

There can be a number of reasons motivating the existence of these investment opportunities, from~which two can be easily identified. The first one can occur when the seller publishes the advertisement in some online listing. In that case, it could occur that the seller ignored the actual value of the house. More likely, the house has remained published in the listing for some time, during which its value has risen (because of global or local trends) but the seller has not updated its value. In the second case, the asset price can be deliberately set lower than its expected value with the intention of it be sold fast, for example if the seller needs the money.

Whichever the case, investment opportunities in real estate exist and can be identified. In this paper, we aim at using machine learning techniques to identify such opportunities, by determining whether the price of an asset is smaller than its estimated value. In particular, we have considered a~dataset of real estate assets located in the Salamanca district of Madrid, Spain, and listed in Idealista, the most relevant Spanish online site for home sales and rentals, during the second semester of 2017.

The remainder of this paper is structured as follows: Section \ref{sec:relwork} describes the state of the art in prediction of houses prices, and places this work in its context. Later, Section \ref{sec:data} describes the dataset used to train the models, with the machine learning techniques being described in Section~\ref{sec:ml}. An~evaluation of the system is performed and its setup and results are discussed in Section \ref{sec:eval}. Finally,~some conclusive remarks and future lines of work are provided in Section \ref{sec:concl}.

%
%
\section{State of the Art}
\label{sec:relwork}

Traditional models used for the appraisal or estimation of real estate assets have relied in hedonic regression, which breaks the asset apart into its constituent characteristics in order to establish a~relationship between each of these and the price of the property. The model learned with hedonic regression can then be used to estimate the price of an asset whose characteristics are known in advance. In recent years, two works using this method have been presented by Jiang et al. \cite{Jiang14,Jiang15} focusing on properties in Singapore with sale transactions between 1995 and 2014. The data used by the authors comprise a total of 315,000 transactions related to 216,000 dwellings in 4,820 buildings, from which some were resales. Properties considered by authors to generate the model are the location, specified~as the postal district, the property type (apartment or condominium) and the ownership type (99 years, 999 years or freehold).

Some related work is concerned with trying to estimate the health of a real estate market using data obtained from search queries and news. In these cases, the housing index price (HPI) is often used as a proxy to determine the status of such market. As an example, Greenstein et al. \cite{Greenstein15} hypothesized that search indices are correlated with the underlying conditions of the United States (US) housing market, and relied on search data obtained from Google Trends as well as housing market indicators for testing that hypothesis. It is remarkable that authors state that traditional forecasting had often relied on statistics provided in annual reports, financial statements, government data, etc., which are often published with significant delay and do not provide enough data resolution. To validate their hypothesis, a~seasonal autoregressive model was used to estimate the relationship. In the end, they are capable of predicting home sales using online search, and also the house price index and demand for home appliances. However, this does not allow the prediction of individual real estate assets. A similar work was also published the same year by Sun et al. \cite{Sun15} who base their research on the fact that 90\% buyers and 92\% sellers rely on the Internet for finding or posting information about properties. In their work, authors~propose combining online daily news sentiments with search engine data from Baidu for predicting the house price index. Prediction models are trained using support vector regression, artificial neural networks trained with back-propagation and radial basis functions neural networks, and tested on data retrieved from Beijing, Shanghai, Chengdu and Hangzhou.

Other works rely on expert systems based on fuzzy logic because of the similarity between this technique and the human approach to decision making \cite{Zurada06}. For example, Guan et al. \cite{Guan14} have used adaptive neuro-fuzzy inference system (ANFIS) for real estate appraisal. To do so, they have used data from 20,192 sales records in a mid-western city of the US between 2003 and 2007, which~were finally reduced to 16,523 after manual curation. The features comprised by this data include location, year~of construction and sale, square footage in each floor (including basement and garage), number~of baths, number of fireplaces, presence of central air, lot type, construction type, wall type and basement type. Their approach combines neural networks with fuzzy inference, and authors report mean average percentage errors for different age bands and different variables sets. It can be seen that ANFIS outperforms multiple regression analysis (MRA), although in some cases using location only instead of all variables can improve results. In 2016, Sarip et al. \cite{Sarip16} presented a work where they used fuzzy least-squares regression FLSR to build a model from 352 properties in the district of Petaling, Kuala~Lumpur. For constructing the model, eight variables were considered: land area, built-up area, number of bedrooms, number of bathrooms, building age, repair condition, quality~of furniture and location, along with the price in Malaysian ringgits (MYR). Those features which were categorial were first converted into numeric attributes to be manageable by the model. The authors~reported a~mean absolute error around 183,000 for FLSR, smaller than ANFIS and artificial neural networks. Finally,~in~2017, del Giudice et al. \cite{Giudice17a} introduced an approach where fuzzy logic was used to evaluate the situations of a real estate market with imprecise information, showing a case study which focused on the purchase of one office building. Interestingly, del Giudice et al. state that, while there is a wide theoretical background regarding real estate investment, empirical support for such theory is scarce in the literature.

A work by Rafiei et al. \cite{Rafiei16} presents an innovative approach where the focus is placed not in property appraisal but in decision making at the time of starting a new construction, i.e., whether ``to~build or not to build", in the words of the authors. Authors describe the use of a deep belief restricted Boltzmann machine for learning a model from 350 condos (3--9 stories) built in Tehran (Iran) between 1993 and 2008. From these condos, a total of 26 attributes are used to train the model. Seven attributes refer to physical and financial properties of the constructions, such as the ZIP code, the total floor area and the lot area, the estimated construction cost per square meter, the duration of construction and the price of the unit at the beginning of the project. The authors also retrieved additional economic variables, which included regulation in the area, prices index, financial status in the construction area, currency exchange rate, population and demographics, etc. When evaluating their model over a test set of another 10 condos, they obtain a test error as low as 3.6\%, a number better than that attained using neural networks trained with backpropagation.

Finally, some authors have relied on the use of machine learning techniques for estimating or predicting the price of individual real estate assets. It is the case of Park and Kwon Bae \cite{Park15}, who~have analyzed housing data of 5359 townhouses in Fairfax County, Virginia, combined from various databases, from 2004 and 2007. These assets had 76 attributes, from which 28 were eventually selected after filtering using a \textit{t}-test and logistic regression. Sixteen of these features are physical features, referring to the number of bedrooms and bathrooms, number of fireplaces, total area, cooling~and heating systems, parking type, etc. In addition, three variables refer to the ratings of elementary, middle~and high schools in the area, and other eight refer to the mortgage contract rate, location~and construction and sale date. The last attribute is the class, which the authors have converted in binary: either the closing price is larger than the listing price or the other way around. Therefore, the problem can be seen as a classification to decide whether an investment is worthy or not instead of a regression problem. For performing classification, authors compare different algorithms: decision trees (C4.5), RIPPER, Naive Bayes and AdaBoost. Best results are achieved using RIPPER (repeated incremental pruning to produce error reduction), a propositional rule learner, with~an~average error of about 25\%. The authors do not provide more specific metrics such as F1 score which allows for determining the model goodness regardless of the class distribution.

Another work was presented by Manganelli et al. \cite{Manganelli16} in 2016, where the focus is put on appraisal based on 148 sales of residential property in a city in the Campania region, Italy. Considered variables are the age of the property, date of sale, internal area, balconies area, connected area, number of services, number of views, maintenance status and floor level. The authors used linear programming to analyze the real estate data, learning the coefficients for the model, achieving an average percentage error of 7.13\%.

Finally, two more works have been published by del Giudice et al. \cite{Giudice17b,Giudice17c} in 2017. In the first of these works, the authors use a genetic algorithm (GA) to try to establish a relationship between rental prices and the location of assets in a central urban of Naples divided into five subareas, considering also the commercial area, maintenance status and floor to be relevant parameters. The way in which the authors use the GA is towards learning weights for these parameters, effectively building a regression model which could be used for estimating rental prices. The obtained absolute average percentage error is equal to 10.62\%. In the second work, the authors rely on Markov chain hybrid Monte Carlo method (MCHMCM), yet they also tested neural networks, multiple regression analysis (MRA) and penalized spline semiparametric method (PSSM). The model is again used for predicting the price of real estate properties in the center of Naples, with the dataset comprising only 65 housing sales in twelve months. Besides the variables from the previous work, they also considered the number of bathrooms, outer surface, panoramic quality, occupancy status and distance from the funicular station. The MCHMCM model achieves an absolute average percentage error of 6.61\%, a better rate that the tested alternatives.

As shown in this section, literature on the application of machine learning to the valuation of real estate assets is relatively scarce, at least in the way we are approaching the problem. While the few last works cited in this section adhere to the same approach than ours, those works are restricted to certain areas and only probes a small set of techniques. In this work, we will also focus on the snapshot of the real estate market in a small region during a period of six months, yet evaluating additional machine learning techniques, and approaching the problem as a regression task.

%
%
\section{Data}
\label{sec:data}

\begin{figure}[H]
\centering
\includegraphics[width=.98\columnwidth]{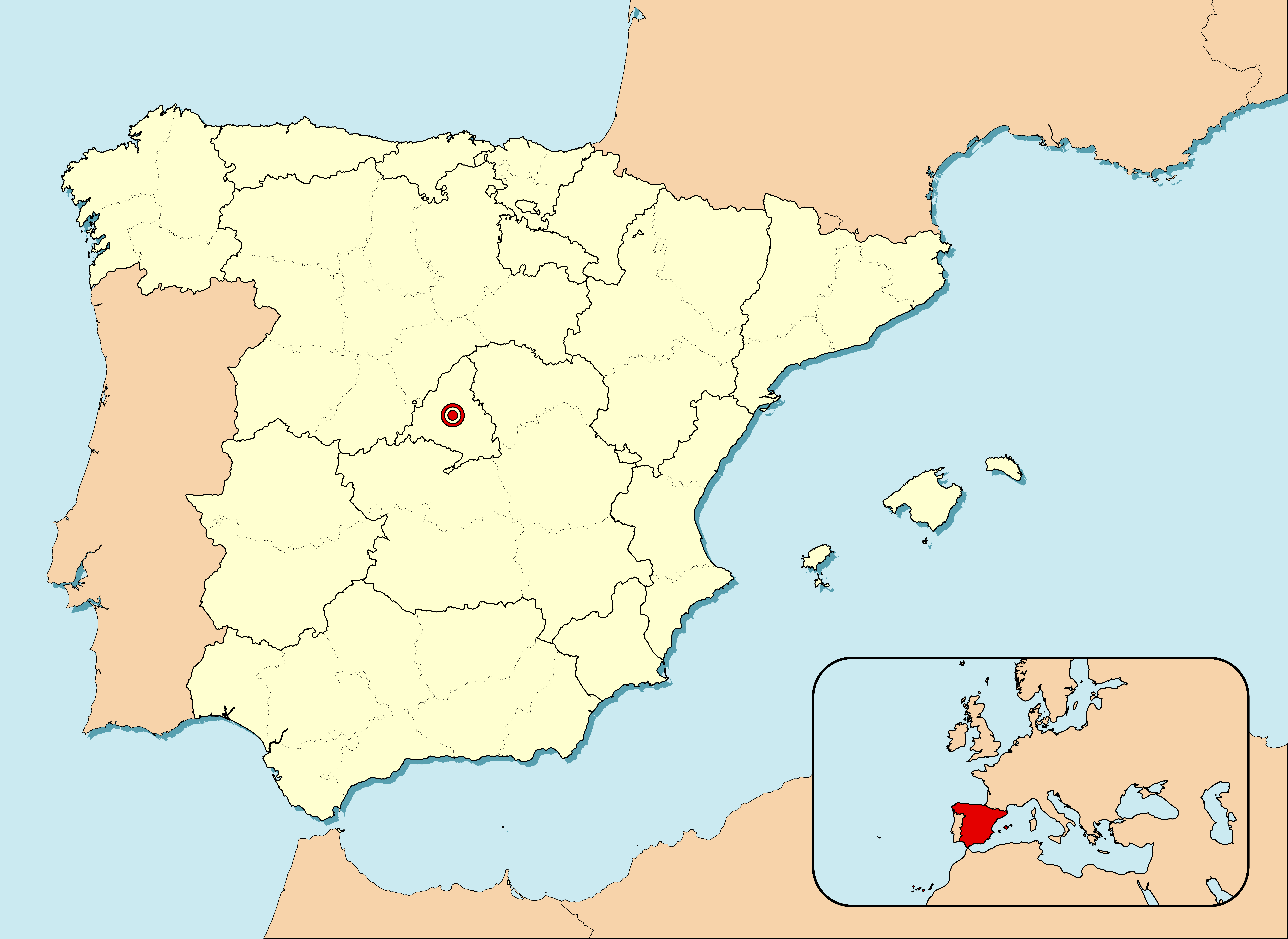}
\caption{Location of the city of Madrid within Spain.}
\label{fig:spain}
\end{figure}

\begin{figure}[H]
\centering
\includegraphics[width=.4\columnwidth]{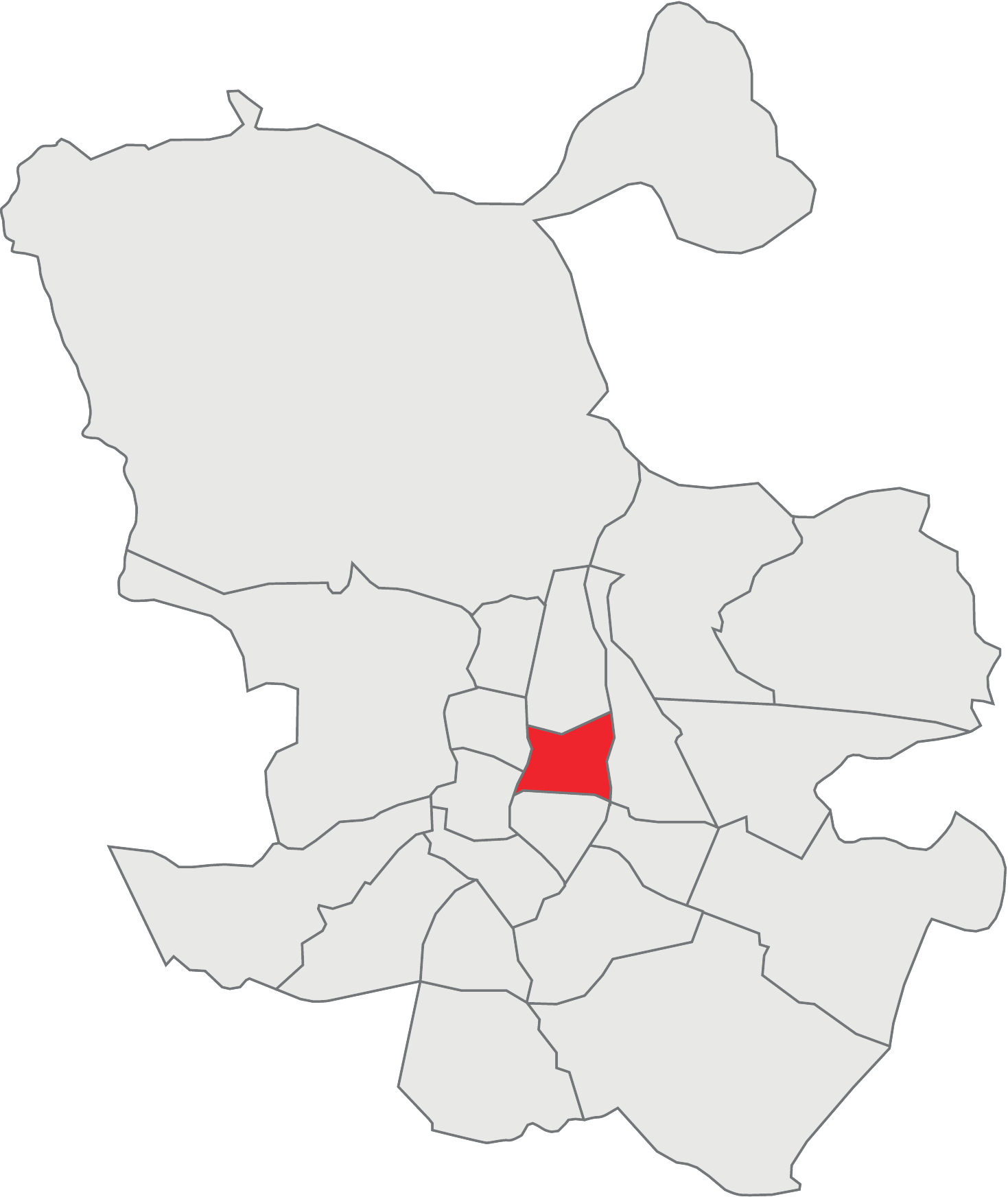}
\caption{Location of the Salamanca district within the city of Madrid.}
\label{fig:salamanca}
\end{figure}

In this paper, we will be focusing in a small segment of the market, comprising high-end real estate assets located in the Salamanca district in Madrid, Spain. The city of Madrid is the capital of Spain and also the largest city, with more than 3 million residents. Its location within Spain (and also framed within the context of the European continent) is shown in Figure \ref{fig:spain}. The city is divided into 21 administrative divisions called districts. The Salamanca district is located to the northeast of the historical center (see Figure \ref{fig:salamanca}) and is home to 150,000 people, being one of the wealthiest and more expensive places in Madrid and Spain.
\unskip

\subsection{Source and Description}

In order to identify opportunities within the framework of a stable market, we have used data from all residences costing more than one million euros, listed in Idealista (the more relevant Spanish online platform for real estate sales and rentals) in the second semester of 2017 (from June 1st to December 31st).

While some features might vary between assets, we consider this sample to be relatively homogeneous and focused in the residential prime market.\ Moreover, some machine learning techniques used in this paper and described in Section \ref{sec:ml} are well suited to deal with a certain degree of heterogeneity, by choosing appropriate features when learning the model based on some measure of discrimination power (e.g., entropy or Gini coefficient).

The raw data used for training the machine learning models comprise several attributes, which~can be grouped in three categories: location, home characteristics and ad characteristics. In particular, the~features involved in determining the location of the real estate asset are the following:
\begin{itemize}
    \item Zone: division within the Salamanca district where the asset is located. This zone is determined by Idealista based on the asset location. 
    \item Postal code: the postal code for the area where the asset is located.
    \item Street name: the name of the street where the asset is located.
    \item Street number: number within the street where the asset is located.
    \item Floor number: the number of the floor where the asset is located.
\end{itemize}

The features defining the asset characteristics are the following:
\begin{itemize}
    \item Type of asset: whether the asset is an apartment (flat, condo...) or a villa (detached or semi-detached house).
    \item Constructed area: the total area of the asset indicated in square meters.
    \item Floor area: the floor area of the asset indicated in square meters, which will in some cases be smaller than the constructed area because terraces, gardens, etc., are ignored in this feature.
    \item Construction year: construction year of the building.
    \item Number of rooms: the total number of rooms in the asset.
    \item Number of baths: the total number of bathrooms in the asset.
    \item Is penthouse: whether the asset is a penthouse.
    \item Is duplex: whether the asset is a duplex, i.e., a house of two floors connected by stairs.
    \item Has lift: whether the building where the asset is located has a lift.
    \item Has box room: whether the asset includes a box room.
    \item Has swimming pool: whether the asset has a swimming pool (either private or in the building).
    \item Has garden: whether the asset has a garden.
    \item Has parking: whether the asset has a parking lot included in the price.
    \item Parking price: if a parking lot is offered by an additional cost, then that price is specified in this~feature.
    \item Community costs: the monthly fee for community costs.
\end{itemize}

Finally, the features defining the ad characteristics are the following:
\begin{itemize}
    \item Date of activation: the date when the ad was published in the listing and displayed publicly.
    \item Date of deactivation: the date when the ad was removed from the listing, which could indicate that it was already sold or rented.
    \item Price: current price of the asset, or price when the ad was deactivated, which in most cases correspond to the price at which the asset was sold or rented.
\end{itemize}

Some of the previous features might not be available. This will happen in most cases because the seller has not explicitly specified some information about the asset, for example, whether a lift is available, which are the monthly community costs, etc. Regarding the street name and number, it~can be intentionally hidden by the seller, so potential buyers are not aware of the actual location of the building. Some information is always available, such as the zone, postal code, constructed area, number of rooms and number of bathrooms.

In the dataset, there is a total of 2266 real estate assets, from which 2174 are apartments and 92~are villas. The asset prices range between 1 and 90 million euros, with an average price of about 2.02~million and a median price of about 1.66 million. More information about the data and the range of values is available in Table \ref{tab:data}.
\begin{table}[H]
\centering
\footnotesize
\setlength{\tabcolsep}{2pt}
\centering
\caption{Data description, showing the range of values for each feature and the number of empty values, as well as mean and standard deviation in the case of numerical values.}
\label{tab:data}
\begin{tabular}{ccccc}
\toprule
\textbf{Feature}    & \textbf{Type}     & \textbf{Range}        & \textbf{Mean (Std. Dev.)}     & \textbf{Empty Values} \\
\midrule
Zone                & Categorical       & 1, 2, 3, 4, 5, 6      & --                            & --                    \\
Postal code         & Categorical       & 28001, 28006, 28009,
                                          28014, 28028, 28046         & --                              & --                  \\
Street name         & Categorical       & 65 values             & --                            & 1453                  \\
Street number       & Categorical       & 77 values             & --                            & 2049                  \\
Floor number        & Categorical       & Basement, Floor,
                                          Mezz, 1--14           & --                            & 119                   \\
Type of asset       & Categorical       & Apartment, Villa      & --                            & --                    \\
Constructed area    & Numerical         & 50--2041 sq.m.       & 288.76 (133.71)               & --                    \\
Floor area          & Numerical         & 93--1700 sq.m.       & 257.63 (126.43)               & 1673                 \\
Construction year   & Numerical         & 1848--2018            & 1953.23 (31.35)               & 1517                 \\
Number of rooms     & Numerical         & 0--20                 & 4.19 (1.35)                   & 6                     \\
Number of baths     & Numerical         & 0--10                 & 3.53 (1.14)                   & 5                     \\
Is penthouse        & Boolean           & T (169)/F (288)     & --                            & 1809                 \\
Is duplex           & Boolean           & T (50)/F (260)      & --                            & 1956                 \\
Has lift            & Boolean           & T (2123)/F (20)    & --                            & 123                   \\
Has box room        & Boolean           & T (1212)/F (269)   & --                            & 785                   \\
Has swimming pool   & Boolean           & T (127)/F (413)     & --                            & 1726                 \\
Has garden          & Boolean           & T (155)/F (391)     & --                            & 1720                 \\
Has parking         & Boolean           & T (687)/F (77)      & --                            & 1502                 \\
Parking price       & Numeric           & 115--750,000          & 52,359.50 (102,670)           & 2209                 \\
Community costs     & Numeric           & 0--3000              & 353.71 (299.61)               & 1536                 \\
\bottomrule
\end{tabular}
\end{table}

\subsection{Data Cleansing}
Before starting to work with the data, we have proceeded to clean some of the data. Although the data is relatively clean, some information is particularly noisy because the way it is communicated by sellers in the listing.

An example is the street name, which is a free text field. For this reason, there was no consensus regarding some aspects such as capitalization, prepositions or accents. We have manually revised this field in order to standardize it according to the actual street names gathered in the streets guide.

The floor area is not specified in most of the cases. In this case, we have assumed it to be equal to the constructed area. Even when this is a rough estimate, it is often the case than the floor area is only slightly smaller than the constructed area, and therefore we have considered this estimation to be an~appropriate proxy.

Finally, for all binary features, we have considered empty values as non-availability of the corresponding features. This seems reasonable within the scope of online listings, since it is more likely that the seller fills the field when the feature is available. For example, in the case of swimming pools, 127 ads specify that the house features a pool, whereas 413 explicitly state that the asset does not include a swimming pool. We will consider that, for those ads where no information is provided about the pool (in this case, a total of 1,726), such a pool is not available.

\subsection{Exploratory Data Analysis}
\begin{figure}[H]
\centering
\includegraphics[width=.8\columnwidth]{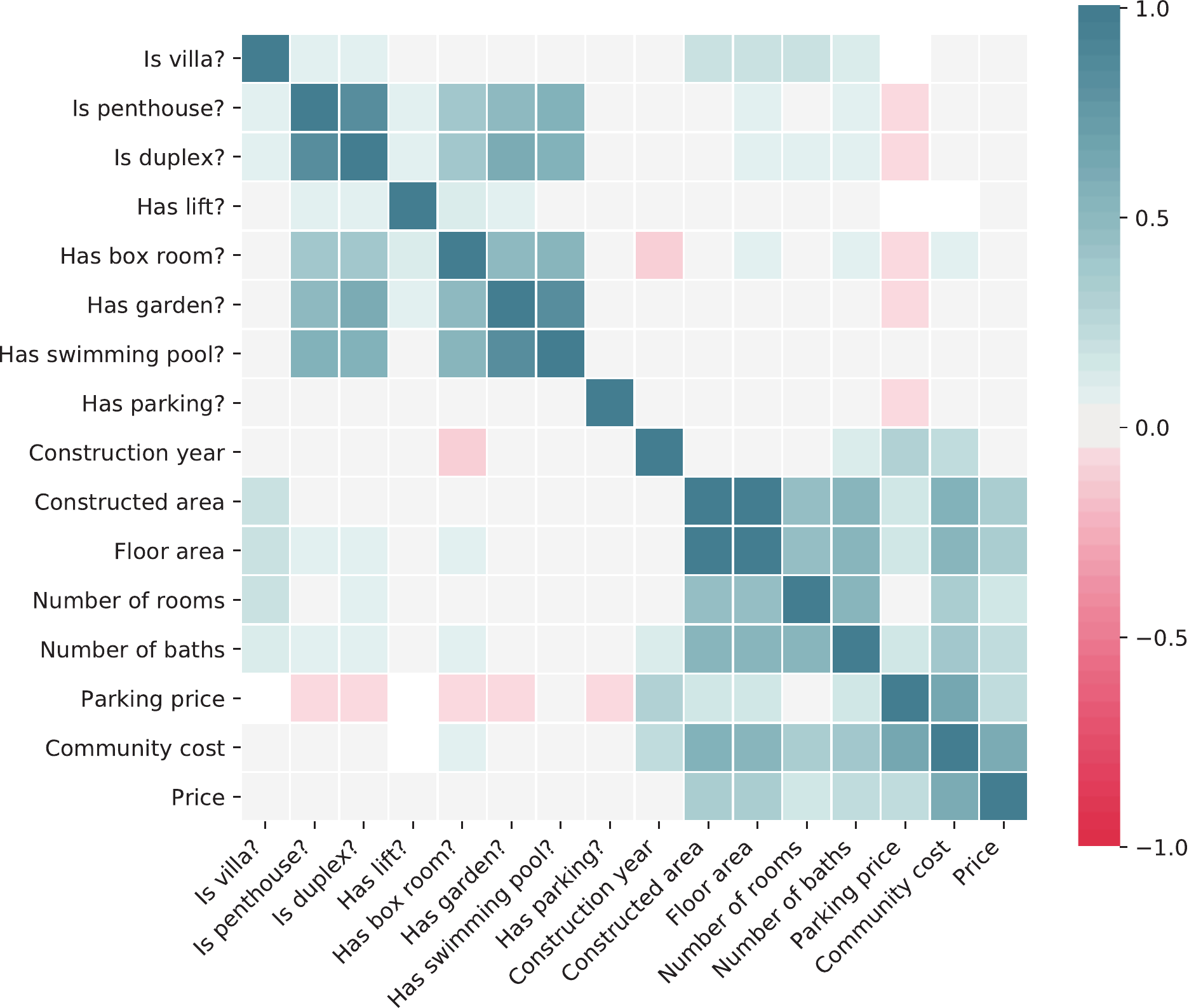}
\caption{Matrix of correlations between each pair of variables.}
\label{fig:corr}
\end{figure}

Before proceeding with the training of machine learning models, we are interested in getting some insights about the data at hand. I\textls[-15]{n particular, it would be interesting to know how different variables affect the price of the real estate model, to understand their potential quality as predictors. This~information is shown in Figure \ref{fig:corr}, where the correlation between each pair of variables (considering~only binary and continuous variables) is shown. The Pearson correlation coefficient is computed for continuous variables, and the point biserial correlation coe}fficient is computed for binary variables.

From the correlation matrix, we can see that the main variables affecting the asset price are those related to the house size, mainly the constructed area. Of course, all of these variables, such as the area, number of bedrooms and number of bathrooms, are highly correlated among them. There is also a~significant impact of the monthly community cost, but, in this case, it is likely that the causality points in the other direction: the cost grows for those houses that are more expensive. Interestingly, most of the binary features as well as the construction year do not seem to have either positive or negative correlation with the price.

\begin{figure}[H]
\centering
\includegraphics[width=.56\columnwidth]{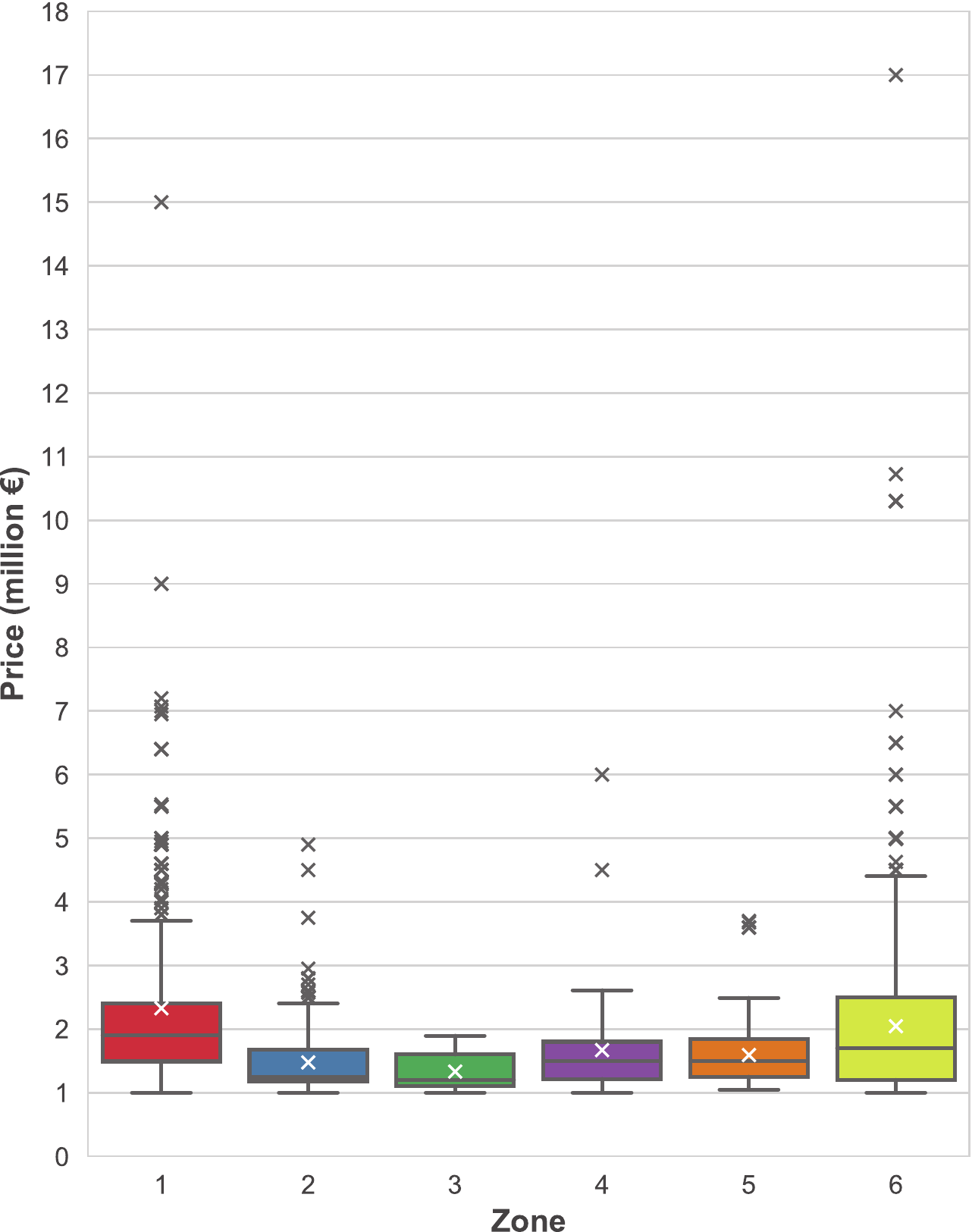}
\caption{Boxplot of the distribution of prices based on the asset location.}
\label{fig:zone_price}
\end{figure}

Figure \ref{fig:zone_price} shows the distribution of asset prices based on their location, more specifically, based on their zone attribute provided by Idealista. Although differences are not very noticeable, it can be seen how assets in zone 1 have a substantially larger mean and median price when compared to the other zones. Zones 2 and 3 have slightly lower prices than the rest of the zones.

\begin{figure}[H]
\centering
\includegraphics[width=.5\columnwidth]{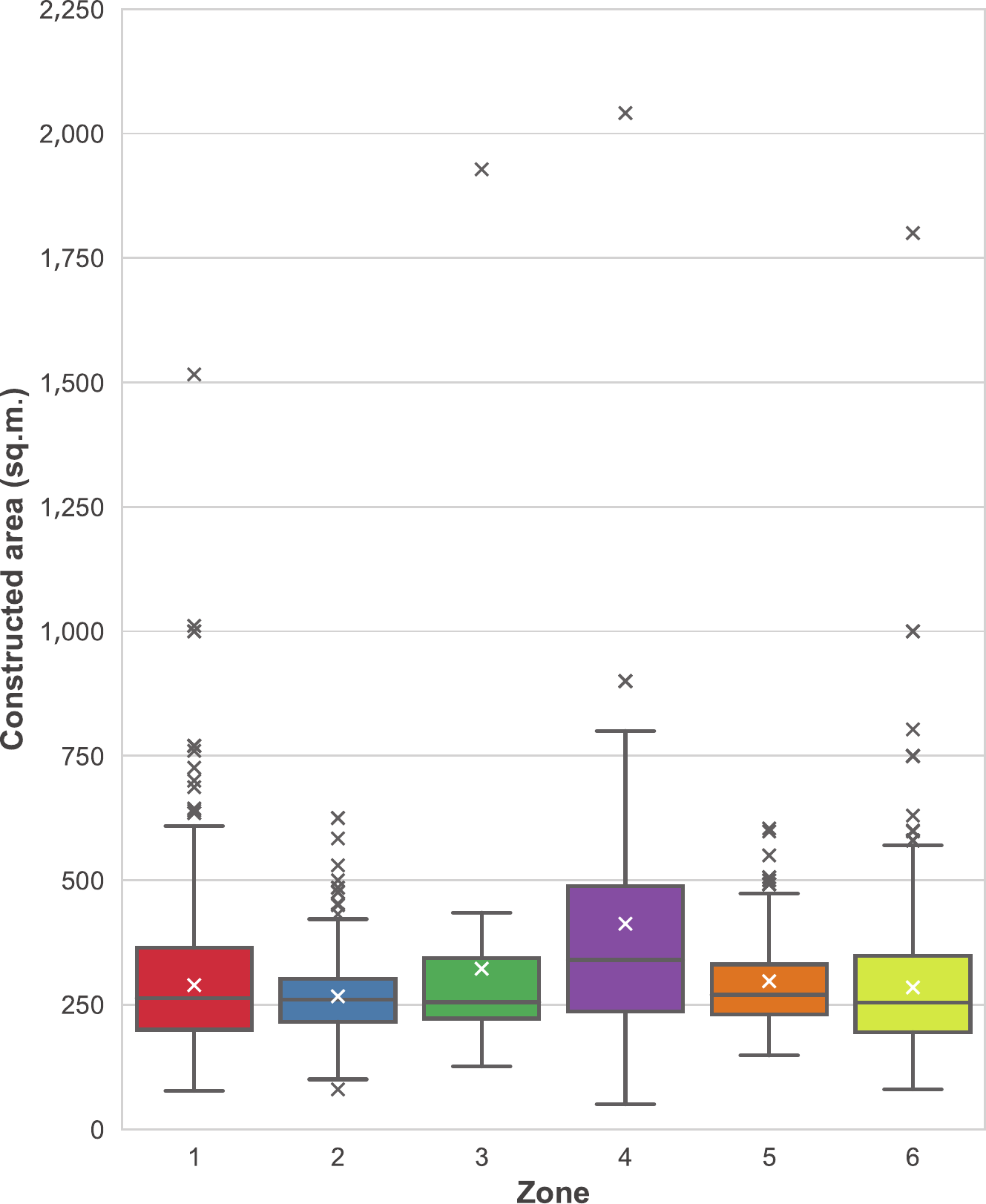}
\caption{Boxplot of the distribution of constructed areas based on the asset location.}
\label{fig:zone_area}
\end{figure}

To understand whether these variations can be explained because of the zone and not because of other factors, we have also plotted the distribution of constructed areas per each zone, which can be seen in Figure \ref{fig:zone_area}. In this figure, we can see how there are not remarkable differences between different zones, with the exception of zone 4 where the median area is noticeably larger. When comparing the distributions of areas with the distribution of prices, we can conclude that the asset location is also a~factor when determining the price, beyond the constructed area itself. For example, it seems that zone 1 is more expensive in average, since prices are larger while areas follow a similar distribution than in other areas. The opposite happens with zone 4, where prices are roughly the same but constructed areas are larger.

\begin{figure}[H]
\centering
\includegraphics[width=.95\textwidth]{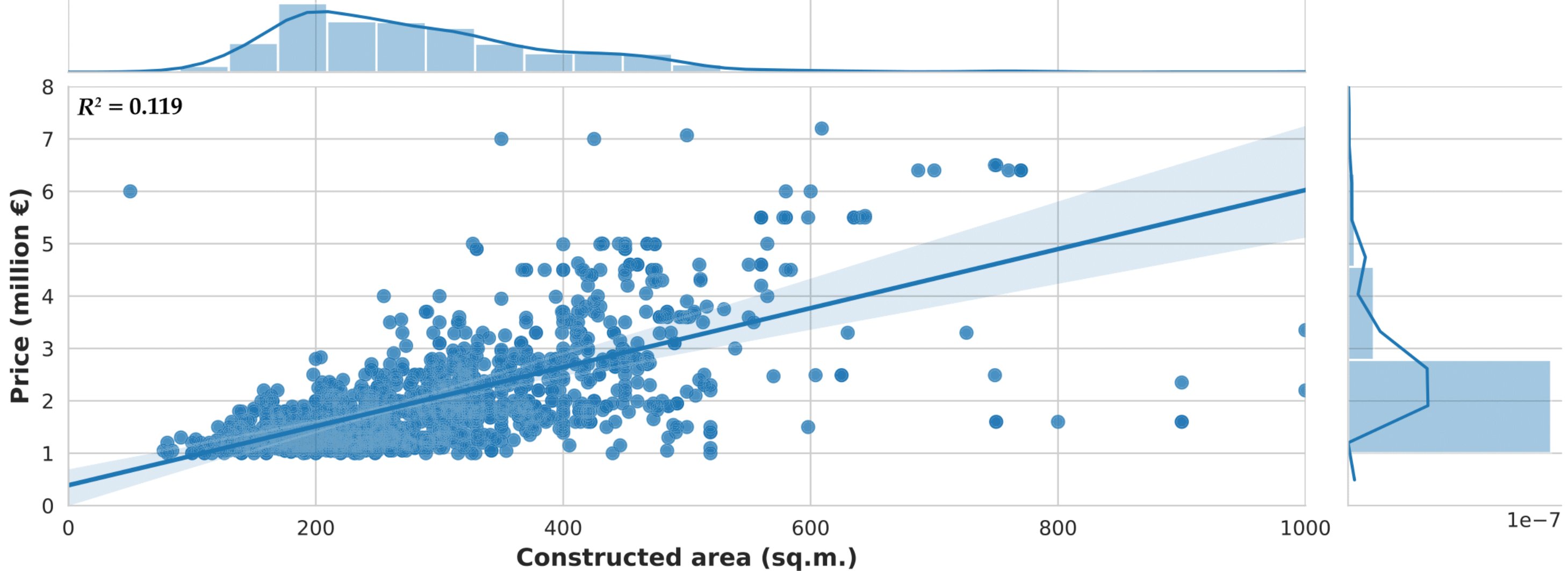}
\caption{Linear regression between the constructed area and the asset price.}
\label{fig:linreg}
\end{figure}

Finally, we are interested in checking how some basic regression models can fit the prices based only on the constructed area. We plot a linear regression model on Figure \ref{fig:linreg}. Assets with prices larger than 8 million have been removed from the figure to ease readability, since these assets are a minority. The coefficient of determination of this linear regression model is $R^2 = 0.119$.

\begin{figure}[H]
\centering
\begin{subfigure}[t]{.48\textwidth}
  \centering
  \includegraphics[width=\textwidth]{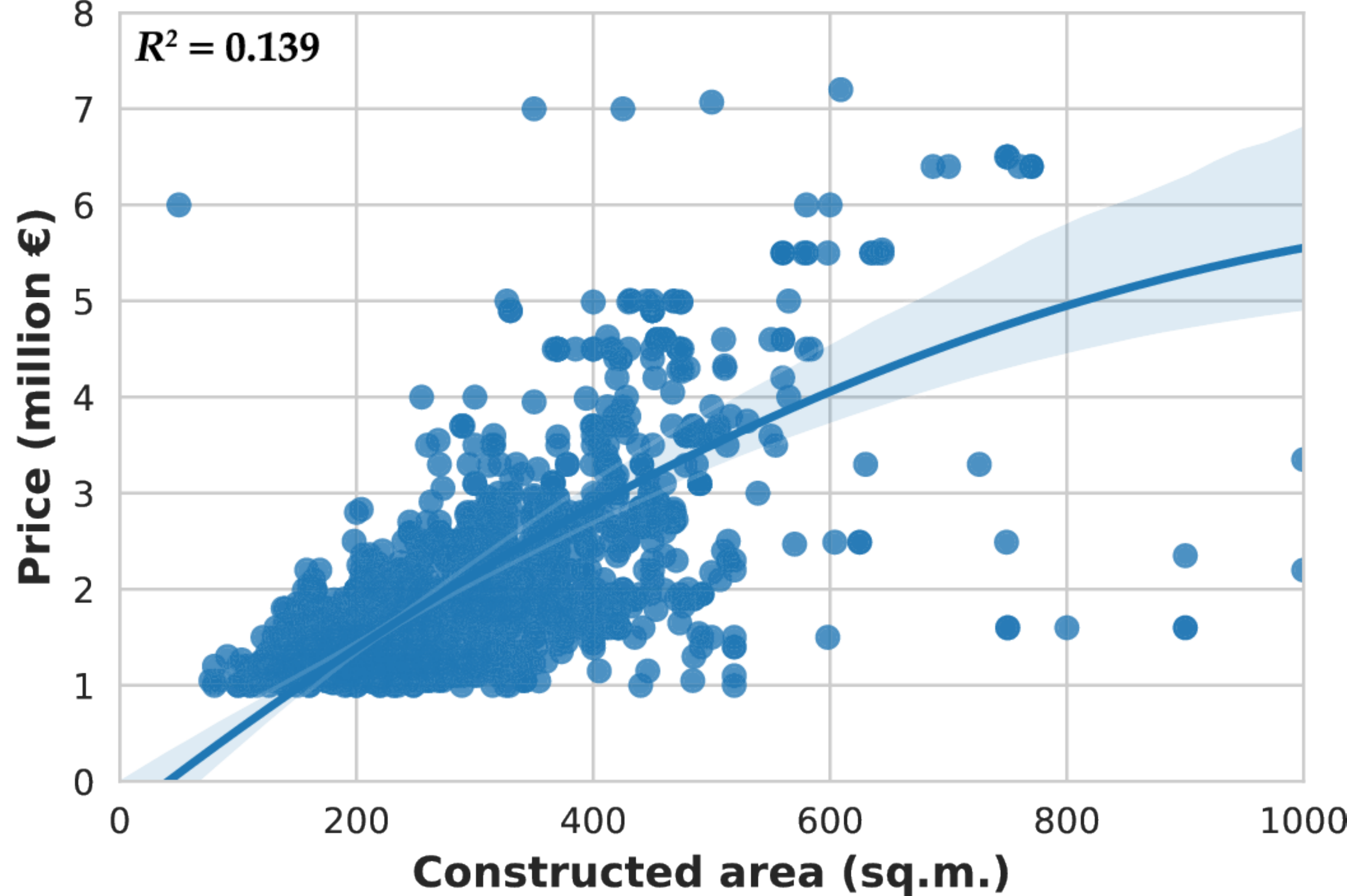}
  \caption{Polynomial of order 2.}
  \label{fig:pol2reg}
\end{subfigure}%
\begin{subfigure}[t]{.48\textwidth}
  \centering
  \includegraphics[width=.95\textwidth]{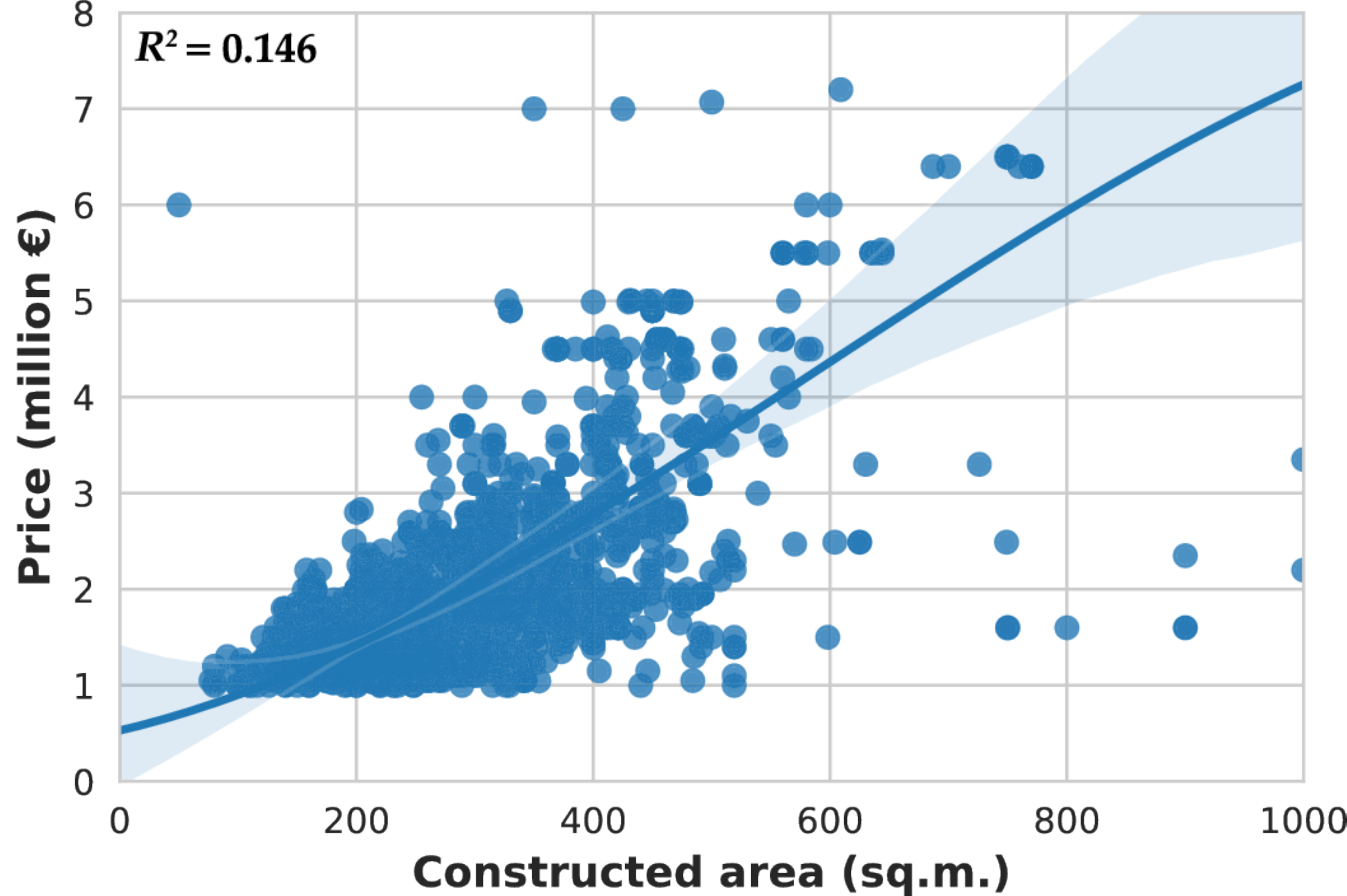}
  \caption{Polynomial of order 3.}
  \label{fig:pol3reg}
\end{subfigure}
\caption{Polynomial regression between the constructed area and the asset price.\label{f8}}
\end{figure}

Additionally, we have also fitted two polynomial regression models of order 2 and 3, shown in Figure \ref{f8}a,b, with respective $R^2$ values of $0.139$ and $0.146$.
\unskip

{We hypothesize} that only the constructed area is insufficient to accurately fit the prices, even when dealing with regression models of higher order than a linear one; and that more complex machine learning algorithm involving a larger set of features could lead to more successful prediction.

To explore the above hypothesis in more detail, we perform two different linear regression models. The first one (“non-robust”), as shown in Table \ref{tab:ols}, represents an ordinary least squares (OLS) estimation in which we consider all variables that will be taken into account for the machine learning models. As we are aware that heteroskedasticity is also a fair concern that clearly could affect these results, we also perform an OLS “robust” estimation following White \cite{white}. Finally, to mitigate existing concerns about the performance of HC0 in small samples, we also explore (unreported) three alternative estimators (HC1,~HC2, and~HC3), following MacKinnon and White \cite{MACKINNON}. Overall our results do not change~qualitatively.
\unskip
\begin{table}[H]
\centering
\caption{Ordinary least squares estimations---in Column 1 ``non-robust" estimation. Column 2 represents a ``robust" estimation (HC0) following White \cite{white}, price in millions. Stars indicate $p$-Value. Legend: * = $p<0.1$, ** = $p<0.5$, *** = $p<0.01$. }
\label{tab:ols}
\begin{tabular}{cccccc}
\toprule
   & \multicolumn{5}{c}                  {\textbf{Price}}    \\\cmidrule{3-4} \cmidrule{5-6}
\textbf{Dependent Var.}                            &                   & \multicolumn{2}{c}{\textbf{Non-Robust}}   & \multicolumn{2}{c}{\textbf{HC0}} \\ \cmidrule{3-4} \cmidrule{5-6}
\textbf{Method: OLS}        &  \textbf{Coef.}   & \textbf{Error}     & \boldmath{$p$}\textbf{-Value}                     & \textbf{Error}     & \boldmath{$p$}\textbf{-Value}         \\
\midrule
Floor number            & 0.045             & 0.023     & *                             & 0.038     & --                  \\
Constructed area        & 0.004             & 0.002     & **                            & 0.001     & ***               \\
Usable area             & 0.001             & 0.002     &  --                             & 0.002     & --                  \\
Is penthouse            & 0.090             & 0.090     &  --                             & 0.090     &  --                 \\
Is duplex               & $-$0.258            & 0.284     & **                            & 0.107     & **                \\
Room number             & $-$0.057            & 0.037     &   --                           & 0.037     & --                  \\
Bath number             & 0.156             & 0.053     & **                            & 0.053     & ***               \\
Has lift                & $-$0.055            & 0.470     &   --                            & 0.756     &  --                 \\
Has box room            & 0.011             & 0.163     &  --                            & 0.061     & --                  \\
Has swimming pool       & 0.050             & 0.243     &                              & 0.095     &  --                 \\
Has garden              & 0.143             & 0.251     &   --                            & 0.079     & *                 \\
Has parking             & 0.260             & 0.244     &  --                             & 0.086     & ***               \\
Is apartment            & $-$0.143            & 0.235     &  --                             & 0.302     & --                  \\
Is villa                & $-$0.065            & 0.278     &    --                           & 0.264     &  --                 \\ \midrule
Location (dummy)        & Yes               &   --        &  --                             &   --        &  --                 \\
Postal code (dummy)     & Yes               &  --          &  --                             &  --         &   --                \\
Street (dummy)          & Yes               &  --         &   --                            &   --        & --                  \\ \midrule
Observations            & 2266             &   --        &   --                            &   --        &   --                \\
\bottomrule
\end{tabular}
\end{table}
%
%
\section{Machine Learning Proposal}
\label{sec:ml}
In the previous section, we have seen how different variables correlate with the asset price, with~the constructed area being the attribute with a larger influence on the price. We have also graphically seen how a linear regression model considering only the constructed area is insufficient for successfully predicting the price. Moreover, in the previous section, we have also shown how several inputs are economically and statistically meaningful explaining house prices with a mutivariate OLS~model.

In this work, we will use machine learning to build and train more complex models for carrying out price prediction. As we explain before, this problem will be tackled as a multi-variate regression problem, with the price being the output, where features can be binary (such as whether there is a~swimming pool or a lift), categorical (such as the asset location) or continuous (such as the constructed area). To be able to use many machine learning techniques, we will convert all categorical features using one-hot encoding. This means that any categorical feature with $N$ values will be transformed into a total of $N$ binary features.

In addition, we know that the values for some of the features remain unknown in some instances. This~happens with the street name, the street number, the construction year, the parking price and the community costs. Since the first feature is categorical, the problem of null values is resolved after performing one-hot encoding (since all binary features can take the value ``false''). As for the other variables, we have decided to remove them from the data. This is especially important for the street number, since it is not a relevant feature but could lead to overfitting (in some cases, the street name and number identify an asset).

Bearing in mind with these constraints, some algorithms are not particularly useful to solve this problem. For example, naive Bayes or Bayesian networks do not often perform well when applied to regression problems \cite{Frank00}. In the case of regression trees, we have considered to not include individual models, since ensembles often perform better.

Because of these reasons, in the end, we have considered the use of four different machine learning techniques, which belong to different categories: kernel models, geometric models, ensembles of rule-based models (regression trees) and neural networks (universal function approximators). The~final set of algorithms is the following:
\begin{itemize}
    \item Support vector regression: this method is also known as a ``kernel method'', and constitutes an extension of classical support vector machine classifiers to support regression \cite{Smola04}. Kernel methods transform data into a higher dimensional state where data can be separable by a certain function, then learning such function to discriminate between instances. When applied to regression, a~parameter epsilon ($\epsilon$) is introduced, thus aiming that the learned function does not deviate from the real output by a value larger than epsilon for each instance.
    \item $k$-nearest neighbors: this is an~example of a geometric technique to perform regression. In this technique, data is not really used to build a model, but is rather considered a ``lazy'' algorithm, since it needs to traverse the whole learning set in order to carry out prediction for a single instance. In particular, what $k$-nearest neighbors does is compute the distance of the instance to be predicted to all of the instances in the learning dataset, based on some distance function or metric (such as Euclidean or cosine distance). Once distances are computed, the $k$ instances in the training set which are closest to the instance subject of prediction will be retrieved, and their known output will be aggregated to generate a predicted output, for example, by computing their average. This method is interesting since it will consider assets similar to the one we want to predict, but can be problematic when dealing with high-dimensional data with binary attributes.
    \item Ensembles of regression trees: regression trees are logical models that are able to learn a set of rules in order to compute the output of an instance given its features. A regression tree can be learned from a training dataset by choosing the most informative feature according to some criterion (such as entropy or statistical dispersion) and then dividing the dataset based on some condition over that feature. The process is repeated with each of the divided training subsets until a tree is completely formed, either because no more features are available to be selected or because fitting a regression model on the subset performs better than computing a regression sub-tree. In this paper, we will use ensembles of trees, which means that several models will be combined in order to reduce regression bias. To build the ensemble, we will use the technique known as extremely randomized trees, introduced by Geurts et al. \cite{Geurts06} as an improvement to random forests.
    \item Multi-layer perceptron: the multi-layer perceptron is an example of a connectionist model, more~particularly an implementation of an artificial neural networks. This kind of model comprises an input layer that receives as input the values for the features of each instance and several hidden layers, each of which features several hidden units, or neurons. The model is fully connected, meaning that each neuron from one layer is connected to every single neuron in the following layer. Each connection has a corresponding floating number, called weight, which serves for aggregating the inputs to the neuron, and then a nonlinear activation function is applied. During training, a gradient descent algorithm is used to fit the connections weights via a~process known as backpropagation.
\end{itemize}

These algorithms can be configured based on different parameters, which can have an impact on performance. Because of the potentially infinite set of possible combinations for these parameters, in~this work, we have decided to test different values only for those that are likely to have a high impact on the outcome. In other cases, we have established default values, or values that are commonly found in the literature. In the following section, we describe the experimental setup along with the different allocations of parameters for testing each of the previously listed machine learning algorithms.

%
%
\section{Evaluation}
\label{sec:eval}
In this section, we will first introduce the experimental setup which has been applied for testing different machine learning regression models. Then, we will describe the results and discuss relevant~findings.

\subsection{Experimental Setup}
As described in the previous section, we will test the performance of four different machine learning techniques. When assessing how these techniques perform, we must be aware that some of them can be configured according to different parameters. In this paper, we have considered the~following:

\begin{itemize}
    \item Support vector regression: we will specify the kernel type.
    \item $k$-nearest neighbors: we will configure the number of neighbors to consider, the distance metric and the weight function used for prediction.
    \item Ensembles of regression trees: we will set up the number of trees that conform the forest, the~criterion for determining the quality of a split and whether or not bootstrap samples are used when building trees.
    \item Multi-layer perceptron: we will consider different architectures, i.e., different configurations of how hidden units are distributed among layers.
\end{itemize}

For training machine learning models, we will use the scikit-learn package for Python \cite{Pedregosa11}, in its version 0.19.2 (the latest as of August 2018). Table \ref{tab:mlsetup} describes the different configurations considered for training the machine learning models. In addition, the canonical names of the classes and the attributes in the scikit-learn package are shown to ease reproducibility of the experiments. In the case of the multi-layer perceptron, each number indicates the number of neurons in one layer, therefore the configuration ``256--128'' means 256 units in the first hidden layer and 128 units in the second hidden~layer.

Additionally, some machine learning techniques perform better when data is normalized. In this case, we will test out all techniques both with data normalized in the range [0, 1] and not normalized. In order to normalize data, we have divided continuous features by the maximum value in the training set. Of course, it could happen that some values in the test are larger than one but that will not constitute an issue. Normalizing data is important for some techniques to work properly: it is the case of neural networks, where big values can lead to the vanishing or exploding gradient problem.

From the four techniques described before, two of them have an stochastic behavior: it is the case of the ensembles of regression trees (where trees in the forest are built based on some random factors) and of the multi-layer perceptron, where weights are initialized randomly. In order to reduce bias and obtain significant results, we have run each of the experiments involving these techniques a total of 30~times.
\newpage
\paperwidth=\pdfpageheight
\paperheight=\pdfpagewidth
\pdfpageheight=\paperheight
\pdfpagewidth=\paperwidth
\newgeometry{layoutwidth=297mm,layoutheight=210 mm, left=2.7cm,right=2.7cm,top=1.8cm,bottom=1.5cm, includehead,includefoot}
\fancyheadoffset[LO,RE]{0cm}
\fancyheadoffset[RO,LE]{0cm}
\begin{table}[H]
\centering
\caption{Configuration in scikit-learn of the different machine learning algorithms that will be used for addressing the regression problem.}
\label{tab:mlsetup}
\begin{tabular}{lll}
\toprule
\textbf{ML algorithm}                   & \textbf{Parameter}                                     & \textbf{Values}  \\
\midrule
Support vector regression               & Kernel type (\texttt{kernel})                          & Radial basis function kernel (\texttt{rbf})       \\
\texttt{svm.SVR}                        & Penalty (\texttt{C})                                   & 1.0                                                  \\
                                        & Kernel coefficient (\texttt{gamma})                    & Inverse of the number of features \\ \midrule
$k$-nearest neighbors                   & Number of neighbors (\texttt{n\_neighbors})            & 5, 10, 20, 50                                     \\
\texttt{neighbors.KNeighborsRegressor}  & Distance metric (\texttt{metric})                      & Minkowski, cosine                              \\
                                        & Weight function (\texttt{weights})                     & Uniform, inverse to distance (\texttt{distance})      \\ \midrule
Ensembles of regression trees           & Number of trees in the forest (\texttt{n\_estimators}) & 10, 20, 50                                        \\
\texttt{ensemble.ExtraTreesRegressor}   & Criterion for split quality (\texttt{criterion})       & Mean absolute error (\texttt{mae}), mean squared error (\texttt{mse})   \\
                                        & Whether bootstrap samples are used (\texttt{bootstrap})& True, false  \\ \midrule
Multi-layer perceptron                  & Network architecture (\texttt{hidden\_layer\_sizes})   & 1024, 256--128, 128--64--32 \\
\texttt{neural\_network.MLPRegressor}   & Activation function (\texttt{activation})              & Rectified linear unit (\texttt{relu}) \\
                                        & Learning rate (\texttt{learning\_rate\_init})          & 0.001 \\
                                        & Optimizer (\texttt{solver})                            & Adam \\
                                        & Batch size (\texttt{batch\_size})                      & 200 \\
\bottomrule
\end{tabular}
\end{table}

\newpage
\restoregeometry
\paperwidth=\pdfpageheight
\paperheight=\pdfpagewidth
\pdfpageheight=\paperheight
\pdfpagewidth=\paperwidth
\headwidth=\textwidth

Finally, in order to prevent biased results when sampling the dataset in order to build the train and test sets, we will use 5-fold cross validation. In this approach, the whole dataset is first randomly shuffled and then splitted into five equally sized folds. Five experiments will then be carried out: one~for each different fold used as the test set. The training set will always be made out of the remaining four folds. When reporting results in the following section, all metrics will refer to the average of the cross validation, i.e., will be computed as the average of the five results obtained for each of the test sets (macro-average).

The total number of experiments that will be carried out is equal to 10 for the support vector regression, 160 for the $k$-nearest neighbors, 3600 for the ensembles of regression trees and 900 for the multi-layer perceptron. It is worth noting that, in all techniques, every combination of setups is run ten times: once per fold both with normalized and not normalized data. In the latter two techniques, each~of these experiments is run 30 times. Therefore, the total number of experiments is 4670.

\subsection{Results and Findings}
In this section, we will explore the performance achieved by different machine learning models. The following quality metrics for regression have been computed, which are provided by the scikit-learn application programming interface (API):

\begin{itemize}
    \item Explained variance regression score, which measures the extent to which a model accounts for the variation of a dataset. Letting $\hat{y}$ be the predicted output and $y$ the actual output, this metric is computed as follows in Equation (\ref{eq:evar}):
    \begin{equation}
    \label{eq:evar}
        E_{var}\left(y, \hat{y}\right) = 1 - \frac{Var\left\{y - \hat{y}\right\}}{Var\left\{y\right\}}.
    \end{equation}

    In this equation, $Var$ is the variance of a distribution. The best possible score is 1.0, which would occur when $y = \hat{y}$.
    \item Mean absolute error, which computes the average of the error for all the instances, computed as follows in Equation (\ref{eq:mae}):
    \begin{equation}
    \label{eq:mae}
        MAE\left(y, \hat{y}\right) = \frac{1}{n} \sum_{i=0}^{n}\left|y_i - \hat{y}_i\right|.
    \end{equation}

    Since this is an error metric, the best possible value is 0.
    \item Median absolute error, similar to the previous score but computing the median of the distribution of differences between the expected and actual values, as shown in Equation (\ref{eq:medae}):
    \begin{equation}
    \label{eq:medae}
        MedAE\left(y, \hat{y}\right) = median\left(\left|y_1 - \hat{y}_1\right|, \dots, \left|y_n - \hat{y}_n\right|\right).
    \end{equation}
    Again, since this is an error metric, the best possible value is 0.
    \item Mean squared error, similar to MAE but with all errors squared, and therefore computed as described in Equation (\ref{eq:mse}):
    \begin{equation}
    \label{eq:mse}
        MSE\left(y, \hat{y}\right) = \frac{1}{n} \sum_{i=0}^{n}\left|y_i - \hat{y}_i\right|^2.
    \end{equation}
    As it happened with MAE, since this is an error metric, the best possible value is 0.
    \item Coefficient of determination ($R^2$), which provides a measure of how well future samples are likely to be predicted. It is computed using Equation (\ref{eq:r2}):
    \begin{equation}
    \label{eq:r2}
        R^2\left(y, \hat{y}\right) = 1 - \frac{\sum_{i=0}^{n}\left(y_i - \hat{y}_i\right)^2}{\sum_{i=0}^{n}\left(y_i - \bar{y}\right)^2}.
    \end{equation}
    In the previous equation, $\bar{y}$ refers to the average of the real outputs. The maximum value for the coefficient of determination is 1.0, which would be obtained when the predicted output matches the real output for all instances. $R^2$ would be 0 if the model always predicts the average output, but it can also hold negative values, since the model can work arbitrarily worse than just predicting the estimated value.
\end{itemize}

\textls[-25]{In order to discuss the results, we will address relevant findings that can be extracted from the~data.}

\subsubsection{Which model performs best?} 

A summary of the results is reported in Table \ref{tab:mlquality}. To build this table, we have averaged the performance for all folds and then averaged over the different experiments (only in stochastic techniques), finally sorting by ascending mean squared error (MSE). Average and standard deviation is reported for those stochastic techniques, whereas the single performance result is shown in the case of deterministic algorithms. To simplify the table, only the best setup for each algorithm is shown in the table. This~is: $k$-nearest neighbors using 50 neighbors, Minkowski distance and a metric inverse to distance; multi-layer perceptron with two layers of 256 and 128 units, respectively; ensembles of regression trees with 50 estimators in the forest, MAE as the split criterion and bootstrapping; and support vector regression with radial basis function (RBF) kernel. These results significantly outperform those reported by a linear regression in all cases.
\begin{table}[H]
\centering
\footnotesize
\renewcommand{\arraystretch}{0.6}
\setlength{\tabcolsep}{2pt}
\caption{Quality metrics per model. Average (and standard deviation) is shown for stochastic models. Only the best setup based on mean MSE is shown. Price in millions.}
\label{tab:mlquality}
\begin{tabular}{llllll}
\toprule
\textbf{ML Algorithm}           & \boldmath{$E_{var}$}    & \boldmath{$MAE$}        & \boldmath{$MedAE$}  & \boldmath{$MSE$}    & \boldmath{$R^2$}  \\
\midrule
$k$-nearest neighbors           & 0.3625 (--)           & 0.4404 (--)           & 0.2068 (--)       & 4.044 (--)        & 0.3598 (--)     \\
Multi-layer perceptron          & 0.3113 (0.0020)       & 0.5637 (0.0026)       & 0.3355 (0.0041)   & 4.2262 (0.0029)   & 0.3067 (0.0027) \\
Ensembles of regression trees   & 0.1303 (0.1381)       & 0.3714 (0.0075)       & 0.1319 (0.0038)   & 4.3468 (0.1548)   & 0.1253 (0.1386) \\
Support vector regression       & 1.73E-5 (--)          & 0.7384 (--)           & 0.4540 (--)       & 4.9015 (--)       & $-$0.0664 (--)    \\
\bottomrule
\end{tabular}
\end{table}

When considering the best performers based on top performance, we notice that ensembles of regression trees work systematically better than other techniques in terms of mean absolute error. In particular, the top ten models are always ensembles, with either 10, 20 or 50 estimators. The~split criterion, bootstrap instances and normalization neither seem to have a significant effect on performance. The smallest mean absolute error is 338,715 euros (a relative error of 16.80\%). Consistent~results are found when considering the median absolute error, although best performers are not sorted on a side-by-side basis (i.e., the model with the best MAE do not correspond to the model with the best median average error (MedAE). In this case, all top-ten models are also ensembles of regression trees. The best median absolute error is 94,850 euros (a relative error of 5.71\%).

\begin{figure}[H]
\centering
\includegraphics[width=.56\columnwidth]{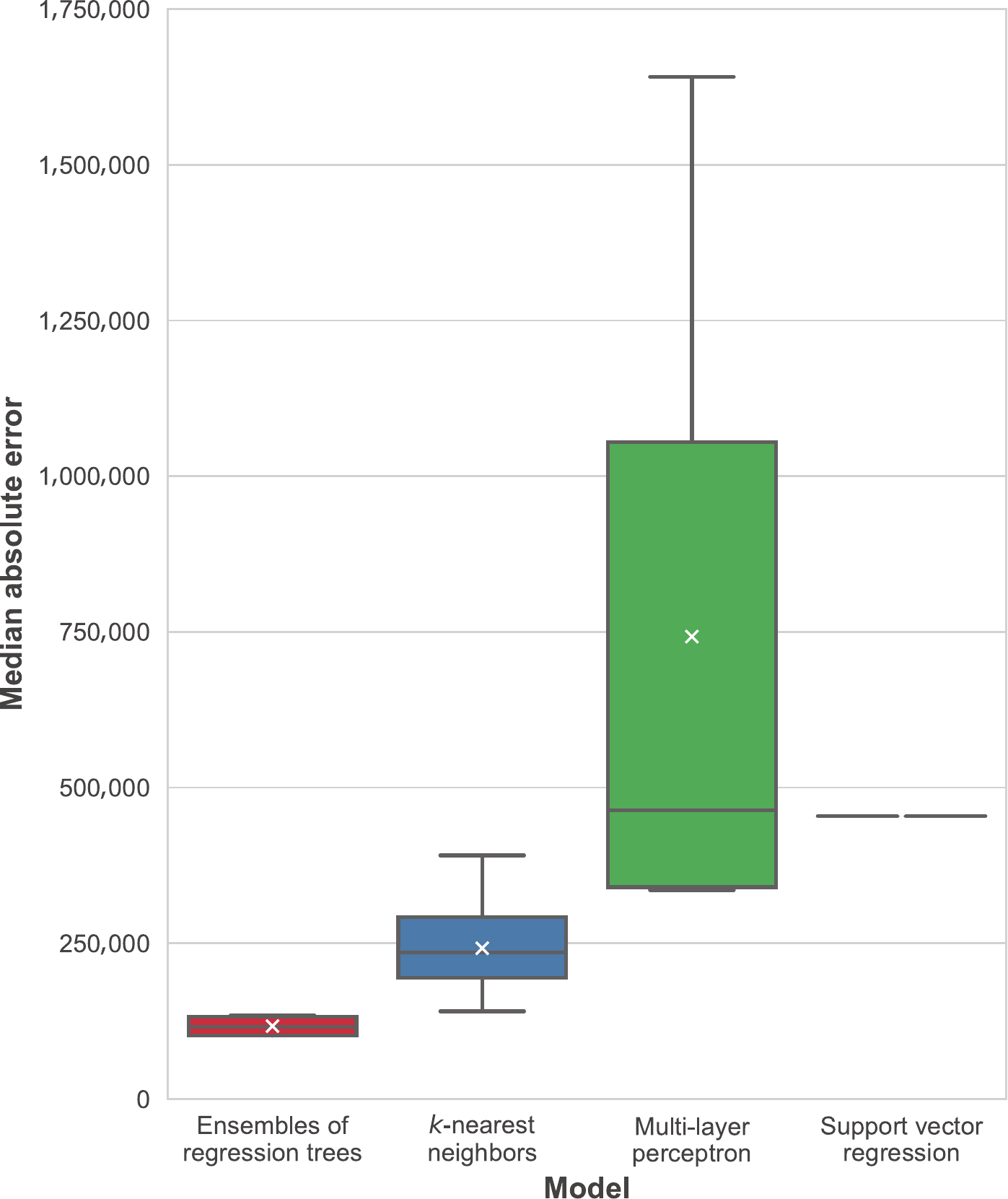}
\caption{Boxplot of the distribution of median average errors per machine learning technique.}
\label{fig:best_medae}
\end{figure}

The distribution of best median average errors per machine learning technique is shown in Figure~\ref{fig:best_medae}. In this figure, we can see how ensembles of regression trees significantly outperform all of the other techniques, followed by $k$-nearest neighbors, support vector regression and the multi-layer perceptron. In the case of support vector regression, only one configuration was tested and the model is deterministic, for this reason there is a single result instead of a distribution. As for the multi-layer perceptron, a very large variance can be seen in the distribution of results, along with a remarkable difference between the average and the median MedAE. Worst MAE and MedAE are always reported by the multilayer-perceptron with a single layer with 1024 units. Because of the relatively small availability of data, this model could be affected by overfitting, although further testing is required to confirm this diagnosis.

Regarding the explained variance regression score and the coefficient of determination, both~metrics are highly correlated. The best model when assessed based on $R^2$ achieves a value of about 0.46 for both metrics.\bigbreak

\subsubsection{How are models affected by setup?} 

\begin{figure}[H]
\centering
\includegraphics[width=.7\columnwidth]{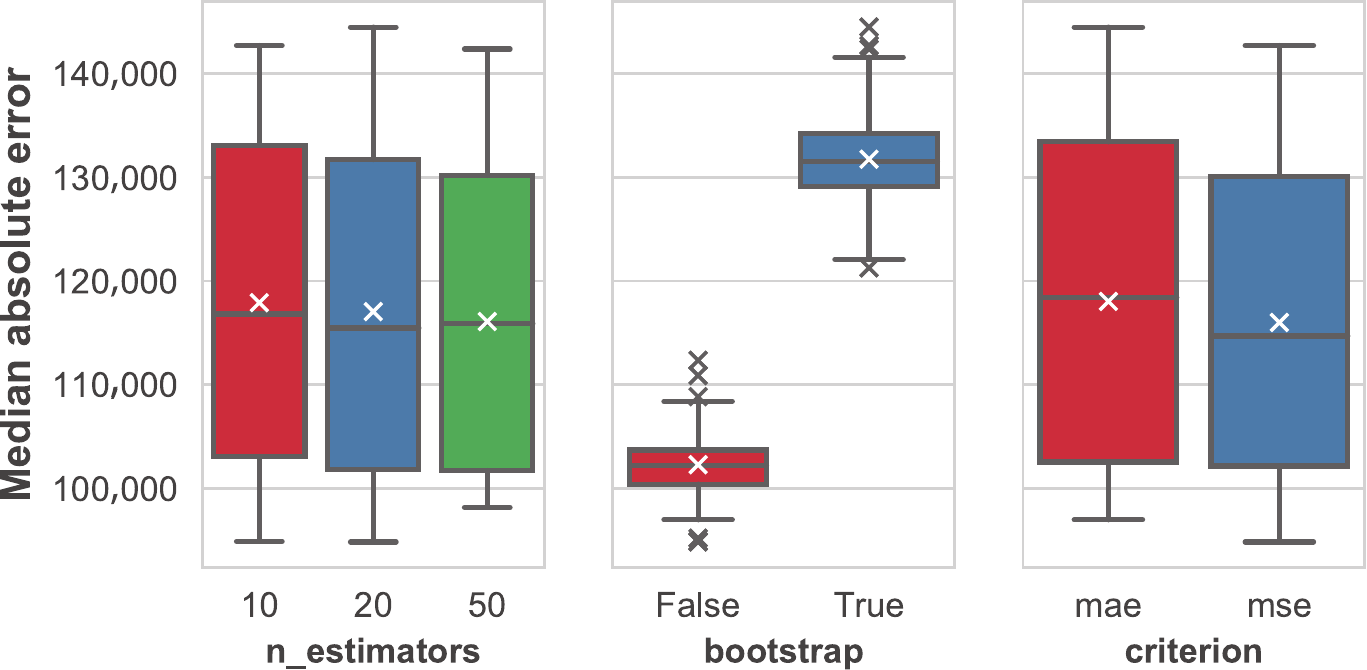}
\caption{Distribution of the median average error based on the different parameters for the ensembles of regression trees.}
\label{fig:setup_extratreesregressor}
\end{figure}

\begin{figure}[H]
\centering
\includegraphics[width=.7\columnwidth]{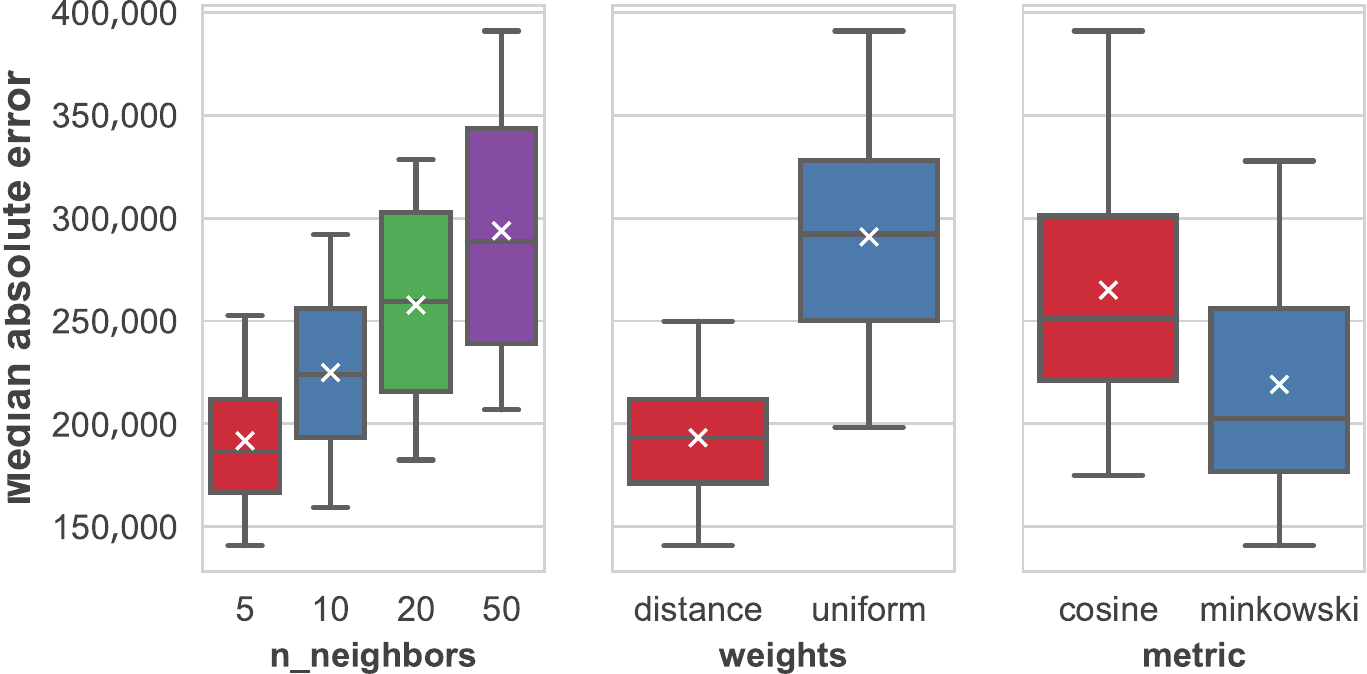}
\caption{{Distribution of the median average error based on the different parameters for the $k$-nearest~neighbors.}}
\label{fig:setup_kneighborsregressor}
\end{figure}

\begin{figure}[H]
\centering
\includegraphics[width=.7\columnwidth]{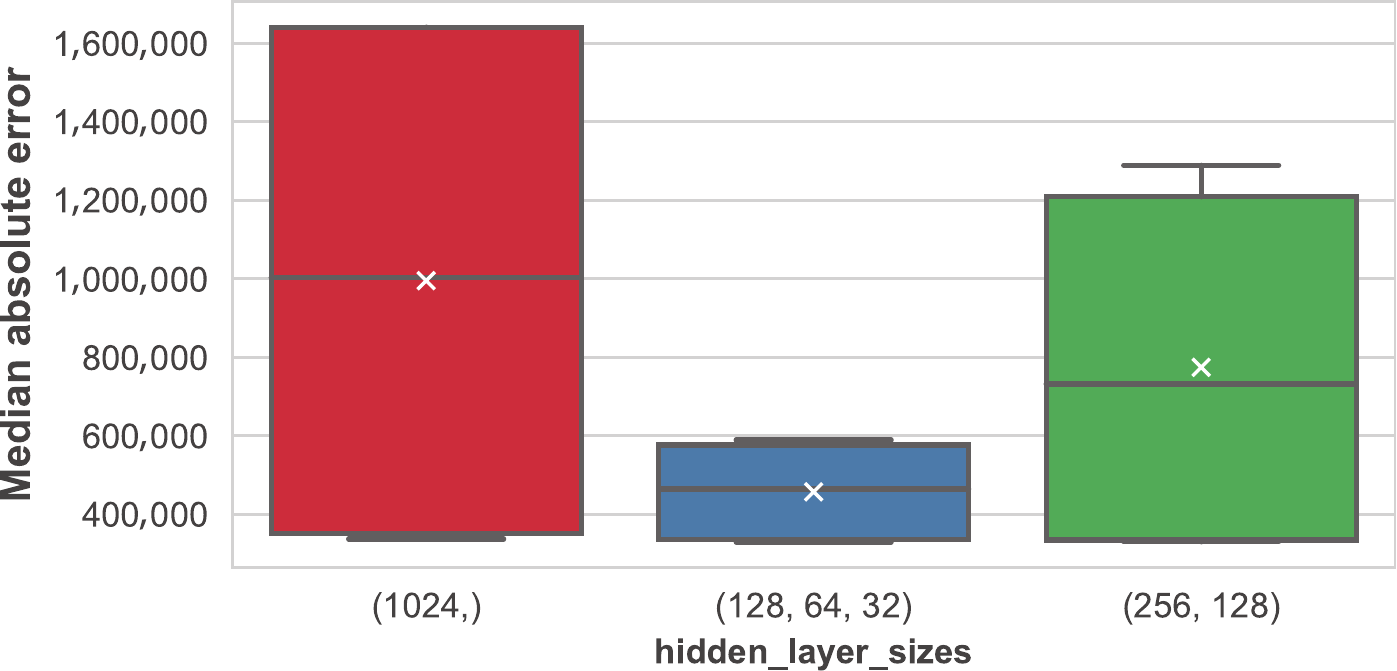}
\caption{Distribution of the median average error based on the different parameters for the multi-layer~perceptron.}
\label{fig:setup_mlpregressor}
\end{figure}

The impact of parameters in the median absolute error for the different machine learning techniques is shown in Figures \ref{fig:setup_extratreesregressor}--\ref{fig:setup_mlpregressor} for the ensembles of regression trees, the $k$-nearest neighbors and the multi-layer perceptron, respectively.

As we can see, in the case of ensembles of regression trees, the number of trees in the forest has little impact on the performance. It seems that the mean MedAE slightly decreases as the number of estimators grow, but the variation is very small. Variance also suffers a reduction as a result of introducing more trees, although again this effect is barely perceptible. In the case of the use of bootstrap sampling, it seems clear that its use negatively affects the model performance in a significant manner. As for the split criterion, a slight reduction in MedAE follows the use of the median squared error (MSE), but, as it happened with the number of estimators, the effect is small.
\unskip
\unskip

When studying the number of neighbors, a clear pattern is shown where the error grows proportionally to the number of considered neighbors. In addition, the variance in MedAE grows as well. This is indicating that considering more similar instances is not only increasing variability but also adding confusion to the regression. From the obtained results, it is left as future work to test this technique with a number of neighbors smaller than five (like one or two) to test whether the error keeps descending. Considering the weights of the different features when computing the nearest neighbors, it is clear that weights that are inversely proportional to the distance of the target instance to the neighbors work better than uniform weights. In addition, regarding the distance metric, Minkowski~has also reported significantly better results than a cosine distance.

Finally, in the case of the multi-layer perceptron, results show that performance is very sensitive to the architecture. It is interesting to see that the minimum MedAE is constant between different alternatives, but the mean and median MedAE is much smaller as more layers with less units each are introduced. As a result, variance is also affected, and we check that the configuration using three hidden layers with 128 units in the first one is the one reporting the best average MedAE. Nevertheless,~as~we concluded earlier, errors are always much larger than those obtained using ensembles of regression trees or $k$-nearest neighbors.

\subsubsection{When does normalization provide an advantage?} 

\begin{figure}[H]
\centering
\includegraphics[width=.7\columnwidth]{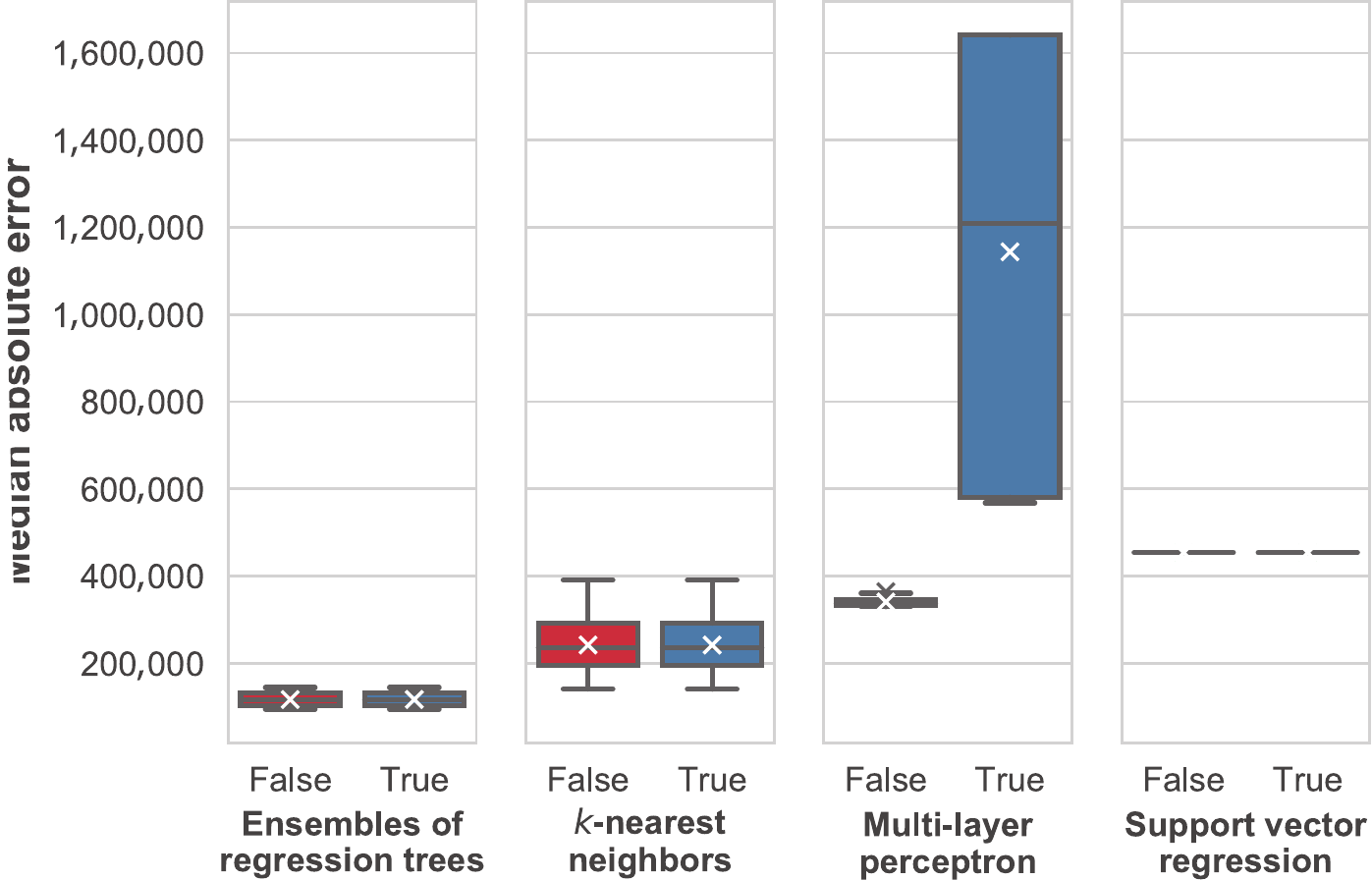}
\caption{Distribution of the median average error based on the use of normalization for the different machine learning techniques.}
\label{fig:norm}
\end{figure}

The impact of the use of normalization in the median average error is shown in Figure \ref{fig:norm} for the different machine learning techniques considered in this paper.

From this figure, we can see how normalization does not have an impact in the ensembles of regression trees, $k$-nearest neighbors and support vector regression. This occurs due to the fact that normalization results in a linear transformation of data. Therefore, in the case of $k$-nearest neighbors, the distance metric is consistent between any pair of instance after the linear transformation of their features. The same thing happens when using support vector machines, and when using such features for building a regression tree.

The only case in which normalization makes a difference is when using a multi-layer perceptron. Interestingly, in this case, the median average error increases when normalization is used. This~is~contrary to the common belief, where normalization is often considered a good practice to prevent numerical instability when training the neural network. In this paper, we will not study the causes of this issue in further detail, since it has been proven that the perceptron is not competitive when compared to other machine learning techniques in the problem of real estate price prediction. Therefore,~a~more thorough study of the causes, involving how the loss value evolves with the training epochs is left for future work.

\subsubsection{How much time do models require to train and run?}  

Training and exploitation times can be a critical aspect of running a machine learning model in a~production setting. For this reason, in this paper, we have decided to introduce a study of how times vary depending on the machine learning technique and its specific setup. It is worth recalling that times reported in this section refer to training and testing within cross-validation setting, where~training is done with 80\% of the instances (1813 assets) and testing with the remaining 20\% (453 assets).

The average time, along with the standard deviation, for both training and prediction of the different machine learning models and configurations is shown in Table \ref{tab:mltimes}. In general, we can see how $k$-nearest neighbors is a lazy machine learning (ML) algorithm, where the training time is negligible when compared with the prediction time, whereas this effect does not happen in other techniques.

For such reason, $k$-nearest neighbors is the fastest algorithm to train, and no significant differences are found for different setups, with the only exception of the distance metric. The reason of Minkowski requiring more time than the cosine distance is that when Minkowski is used, some pre-processing can be done during training in order to accelerate prediction, such as building a KD-tree ($k$-dimensional tree) or a Ball-tree. As~a result, training is faster when this pre-processing is not done (because it is not possible using a~cosine distance) but prediction is slower.

Support vector regression's training is relatively fast when compared to the neural network models or ensembles of trees, but prediction time is about two orders of magnitude higher.

In the case of ensembles, it can be seen how the setup has a remarkable effect on both training and prediction time. These times scale linearly with the number of trees, which is an expected behavior. The use of bootstrap instances and the split criterion do not affect the prediction time, but~impact significantly the cost of training. This cost is especially sensitive to the training split criterion, where~computing the MAE requires about 50 times the time of computing the MSE. This is likely to be due to implementation details of the error computation.
\begin{table}[H]
\centering
\footnotesize
\renewcommand{\arraystretch}{0.6}
\setlength{\tabcolsep}{2pt}
\caption{Average (and standard deviation) of training and prediction times for different machine learning configurations.}
\label{tab:mltimes}
\begin{tabular}{lllll}
\toprule
\textbf{ML Algorithm}           & \textbf{Parameter}    & \textbf{Value}    & \textbf{Train Time (s)}   & \textbf{Predict Time (s)}  \\
\midrule
Support vector regression       & --                    & --                & 0.500 (0.022)             & 0.093 (0.0015)      \\
\midrule
\multirow{ 8}{*}{Ensembles of regression trees}
                                & \multirow{ 3}{*}{\texttt{n\_estimators}}
                                                        & \texttt{10}       & 0.999 (1.068)             & 0.0011 (0.00006) \\
                                &                       & \texttt{20}       & 2.037 (2.185)             & 0.0021 (0.00011) \\
                                &                       & \texttt{50}       & 5.192 (5.587)             & 0.0049 (0.00024) \\ \cmidrule{2-5}
                                & \multirow{ 2}{*}{\texttt{bootstrap}}
                                                        & \texttt{true}     & 1.752 (2.250)             & 0.0026 (0.0015) \\
                                &                       & \texttt{false}    & 3.734 (4.905)             & 0.0028 (0.0017) \\ \cmidrule{2-5}
                                & \multirow{ 2}{*}{\texttt{criterion}}
                                                        & \texttt{mae}      & 5.337 (4.195)             & 0.0027 (0.0016) \\
                                &                       & \texttt{mse}      & 0.149 (0.103)             & 0.0027 (0.0016) \\ \midrule
\multirow{ 10}{*}{$k$-nearest neighbors}
                                & \multirow{ 4}{*}{\texttt{n\_neighbors}}
                                                        & \texttt{5}        & 0.0016 (0.0014)           & 0.013  (0.0038) \\
                                &                       & \texttt{10}       & 0.0017 (0.0015)           & 0.014  (0.0036) \\
                                &                       & \texttt{20}       & 0.0017 (0.0016)           & 0.015  (0.0031) \\
                                &                       & \texttt{50}       & 0.0016 (0.0015)           & 0.018  (0.0016) \\ \cmidrule{2-5}
                                & \multirow{ 2}{*}{\texttt{weights}}
                                                        & \texttt{distance} & 0.0017 (0.0015)           & 0.015  (0.0035) \\
                                &                       & \texttt{uniform}  & 0.0016 (0.0014)           & 0.015  (0.0037) \\ \cmidrule{2-5}
                                & \multirow{ 2}{*}{\texttt{metric}}
                                                        & \texttt{cosine}   & 0.00027 (0.000012)        & 0.018  (0.0013) \\
                                &                       & \texttt{minkowski}& 0.0030 (0.00038)          & 0.012  (0.0031) \\ \midrule
\multirow{ 3}{*}{Multi-layer perceptron}
                                & \multirow{ 3}{*}{\texttt{hidden\_layer\_sizes}}
                                                        & \texttt{(128, 64, 32)} & 1.727 (1.104)        & 0.00076 (0.00013) \\
                                &                       & \texttt{(256, 128)}    & 3.045 (1.487)        & 0.0011 (0.000061) \\
                                &                       & \texttt{(1024)}        & 8.376 (0.094)        & 0.0021 (0.000061) \\
\bottomrule
\end{tabular}
\end{table}

Finally, in the case of the multi-layer perceptron, times are in a similar scale than those of the ensembles. Interestingly, the time in this case decreases as more layers are added, given that the number of units is smaller. This is likely to be due to the implementation, although a deeper study is not within the scope of this paper and is left for future work.

%
%
\section{Conclusions}
\label{sec:concl}
The real estate market constitutes a good setting for investing, due to the many aspects governing the prices of real estate assets and the variances that can be found when looking at local markets and small-scale niches.

In this paper, we have explored the application of diverse machine learning techniques with the objective of identifying real estate opportunities for investment. In particular, we have focused first on the problem of predicting the price of a real estate asset whose features are known, and have modeled it as a regression problem.

We have performed a thorough cleansing and exploration of the input data, after which we have decided to build machine learning models using four different techniques: ensembles of regression trees, $k$-nearest neighbors, support vector machines for regression and multi-layer perceptrons. Cross-validation of five folds have been used in order to avoid biases resulting from the split in train and test subsets. Because we understand that the parameterization of the different techniques can drive significant variations in the performance, we have identified some of the potentially most influencing parameters and tested different setups for those. We have also reported results on the use of the algorithms after data normalization.

After training and evaluating the models, we have thoroughly studied the results, revealing~findings on how different setups impact the performance. Results prove that outperforming models are always those consisting of ensembles of regression trees. In quantitative terms, we have found that the smallest mean absolute error is 338,715 euros, and the best median absolute error is 94,850 euros. These errors can be considered high within the scope of financial investment, but are relatively small under the fact that data comprise only assets with values over one million euros. In~fact, when the mean and median absolute errors are compared with the mean and median of the distribution of prices, relative errors of 16.80\% and 5.71\% are obtained, respectively. These errors are significantly smaller than the ones provided by a classical linear regression model, therefore highlighting the advantage of more complex machine learning algorithms.

In this sense, it is worth mentioning that the fact of the mean being much larger than the median can be explained due to the presence of outliers. For example, the most expensive asset in the dataset is priced 90 million dollars, and, according to the description, it is an apartment of 473 square meters with five bedrooms and five bathrooms. This price seems to be excessive for an apartment of such characteristics, meaning that either the price is a typo and the asset was sold at a much smaller price, or the description has been recorded incorrectly.

In either case, further analysis of prediction errors, even involving manual assessment of experts, is left for future work which can serve for improving the quality of the database and therefore of the trained machine learning models.

In addition, the evaluation of a model built using $k$-nearest neighbors with a number of neighbors smaller than five was left for future work. Additionally, further analysis on the impact of normalization in the performance of the multi-layer perceptron can be of interest, since results reported in this paper seems to contradict the common belief that normalization helps prevent numerical instability and can ease faster convergence. Finally, the impact of different architectures in both regression error and time are worth exploring.

Another field for potential research involves the use of deep learning techniques for extracting relevant features from natural language descriptions. So far, these data have not been provided to us, but all adverts have a description introduced by the seller describing the home at sale. A feature vector extracted from this text, for example by using convolutional neural networks with temporal components, could add a remarkable value to the features already known. Relevant deep learning techniques for this purpose have been recently surveyed by Liu et al. \cite{Liu17}.

Another potential future research line lies in the use of classification techniques in order to differentiate between market segments. In other words, in a bigger sample, the coexistence, within the same sample, of several market segments could derive in unreliable estimates if they are completely neglected. On those cases, a two-step process would be appropriate in which, first, we segment the sample and, finally, we apply the regression models as proposed in this work.

Additionally, in this paper, we have approached the problem of identifying investment opportunities as a regression problem consisting of the estimation of the actual appraisal of the assets. However, if this valuation were done manually, then the problem could be tackled as a~binary classification problem, where the objective would be to determine whether the asset itself is an~investment opportunity; for example, if the sale price were smaller than the valuation price.

Finally, in this work, we have constrained to the analysis of a snapshot of the real estate market in a six-month period. However, it could be interesting to consider the modeling of time series for prediction, since, in some cases, it has been proved that temporal information can improve substantially the prediction performance \cite{Deters17}. Exploring this research line is also suggested as a future work.

%
%
\section{{Data Statement}}  
This work has relied on data provided by Idealista, consisting of the list of real estate assets for sale in the platform in the Salamanca district (Madrid, Spain) during the second semester of 2017 (between July 1st and December 31st, both inclusive). Due to our non-disclosure agreement with Idealista, we are not allowed to publish or redistribute the dataset used for this work. For authors aiming at reproducing experiments in this paper, we suggest contacting Idealista using the form available in the following address: \url{https://www.idealista.com/data/empresa/sobre-nosotros}.

\vspace{12pt}
%
%
\authorcontributions{Conceptualization, A.B., I.B., A.J.M.; Data curation, A.B.; Formal analysis, A.B.; Investigation, A.B.; Methodology, A.B., R.I.; Validation, A.B., I.B., A.J.M., R.I., \'O.B., C.A.; Writing--original draft, A.B., Writing--reviewer changes, A.B., R.I.; Writing--review \& editing, A.B., I.B., A.J.M., R.I., \'O.B., C.A.; Funding acquisition, I.B., A.J.M.; Project administration; I.B., A.J.M.; Resources, I.B.; Supervision, I.B., A.J.M., \'O.B.} 

\conflictsofinterest{
Some of the knowledge obtained from this paper or the machine learning models built in this work might be implemented and/or commercialized in the future by Rentier Token (Madrid, Spain), a company to which all authors in the present work are affiliated. The authors declare no other conflicts of interest.}
\funding{This research received no external funding.}
%
%
\acknowledgments{The authors would like to thank Beatriz C\'{a}mara and Alfonso Lozano from Idealista for their collaboration in providing access to the data used in this paper and support; and also David Quintana for his suggestions and insights to improve the final version of the paper.}


\reftitle{References}





\begin{thebibliography}{999}
\providecommand{\natexlab}[1]{#1}

\bibitem[Teuben and Bothra(2018)]{Teuben18}
Teuben, B.; Bothra, H.
\newblock {\emph{Real Estate Market Size 2017---Annual Update on the Size of the
  Professionally Managed Global Real Estate Investment Market}};
\newblock Technical Report; MSCI, Inc.: New York City, NY, USA, 2018.
\newblock Available online:
  \url{https://www.msci.com/documents/10199/6fdca931-3405-1073-e7fa-1672aa66f4c2} (accessed on 20 November 2018).   

\bibitem[Idealista(2018)]{Idealista18}
Idealista.
\newblock {\'{I}ndice Idealista 50: Evoluci\'{o}n del Precio de la Vivienda de
  Segunda Mano en Espa\~{n}a},  2018.
\newblock Available online:
  \url{https://www.idealista.com/news/estadisticas/indicevivienda#precio} (accessed on 20 November 2018).   

\bibitem[Jiang \em{et~al.}(2014)Jiang, Phillips, and Yu]{Jiang14}
Jiang, L.; Phillips, P.C.B.; Yu, J.
\newblock {A New Hedonic Regression for Real Estate Prices Applied to the
  Singapore Residential Market}.
\newblock Technical Report, Cowles Foundation Discussion Paper No. 1969,  2014.
\newblock Available online: \url{https://ssrn.com/abstract=2533017} (accessed on 20 November 2018).   

\bibitem[Jiang \em{et~al.}(2015)Jiang, Phillips, and Yu]{Jiang15}
Jiang, L.; Phillips, P.C.; Yu, J.
\newblock {New Methodology for Constructing Real Estate Price Indices Applied
  to the Singapore Residential Market}.
\newblock {\em J. Bank. Financ.} {\bf 2015}, {\em
  61},~S121--S131.

\bibitem[Greenstein \em{et~al.}(2015)Greenstein, Tucker, Wu, and
  Brynjolfsson]{Greenstein15}
Greenstein, S.M.; Tucker, C.E.; Wu, L.; Brynjolfsson, E.
\newblock {The Future of Prediction : How Google Searches Foreshadow Housing
  Prices and Sales The Future of Prediction How Google Searches Foreshadow
  Housing Prices}. In {\em Economic Analysis of the Digital Economy}; The
  University of Chicago Press: Chicago, IL, USA, 2015; pp. 89--118.

\bibitem[Sun \em{et~al.}(2015)Sun, Du, Xu, Zuo, Zhang, and Zhou]{Sun15}
Sun, D.; Du, Y.; Xu, W.; Zuo, M.; Zhang, C.; Zhou, J.
\newblock {Combining Online News Articles and Web Search to Predict the
  Fluctuation of Real Estate Market in Big Data Context}.
\newblock {\em Pac. Asia J. Assoc. Inf. Syst.}
  {\bf 2015}, {\em 6},~19--37.

\bibitem[Zurada \em{et~al.}(2016)Zurada, Levitan, and Juan]{Zurada06}
Zurada, J.; Levitan, A.; Juan, G.
\newblock {Non-Conventional Approaches to Property Vale Assessment}.
\newblock {\em J. Appl. Bus. Res.} {\bf 2016}, {\em
  22},~1--14.

\bibitem[Guan \em{et~al.}(2014)Guan, Shi, Zurada, and Levitan]{Guan14}
Guan, J.; Shi, D.; Zurada, J.M.; Levitan, A.S.
\newblock {Analyzing Massive Data Sets: An Adaptive Fuzzy Neural Approach for
  Prediction, with a Real Estate Illustration}.
\newblock {\em J. Organ. Comput. Electron. Commer.}
  {\bf 2014}, {\em 24},~94--112.

\bibitem[Sarip \em{et~al.}(2016)Sarip, Hafez, and Nasir~Daud]{Sarip16}
Sarip, A.G.; Hafez, M.B.; Nasir~Daud, M.
\newblock {Application of Fuzzy Regression Model for Real Estate Price
  Prediction}.
\newblock {\em Malays. J. Comput. Sci.} {\bf 2016}, {\em
  29},~15--27.

\bibitem[Del~Giudice \em{et~al.}(2017)Del~Giudice, De~Paola, and
  Cantisani]{Giudice17a}
Del~Giudice, V.; De~Paola, P.; Cantisani, G.
\newblock {Valuation of Real Estate Investments through Fuzzy Logic}.
\newblock {\em Buildings} {\bf 2017}, {\em 7},~26.

\bibitem[Rafiei and Adeli(2016)]{Rafiei16}
Rafiei, M.H.; Adeli, H.
\newblock {A Novel Machine Learning Model for Estimation of Sale Prices of Real
  Estate Units}.
\newblock {\em J. Construct. Eng. Manag.} {\bf 2016},
  {\em 142},~04015066.

\bibitem[Park and Kwon~Bae(2015)]{Park15}
Park, B.; Kwon~Bae, J.
\newblock {Using Machine Learning Algorithms for Housing Price Prediction: The
  Case of Fairfax County, Virginia Housing Data}.
\newblock {\em Expert Syst. Appl.} {\bf 2015}, {\em
  42},~2928--2934.

\bibitem[Manganelli \em{et~al.}(2016)Manganelli, Paola, and
  Giudice]{Manganelli16}
Manganelli, B.; Paola, P.D.; Giudice, V.D.
\newblock {Linear Programming in a Multi-Criteria Model for Real Estate
  Appraisal}.
\newblock In Proceedings of the  International Conference on Computational Science and its
  Applications, Part I, Salamanca, Spain,  12--16 November 2007; Volume {9786}, pp.  182--192.

\bibitem[Del~Giudice \em{et~al.}(2017{\natexlab{a}})Del~Giudice, De~Paola, and
  Forte]{Giudice17b}
\textls[-25]{Del~Giudice, V.; De~Paola, P.; Forte, F.
\newblock {Using Genetic Algorithms for Real Estate Appraisals}.
\newblock {\em Buildings} {\bf 2017}, {\em 7},~31.}

\bibitem[Del~Giudice \em{et~al.}(2017{\natexlab{b}})Del~Giudice, De~Paola,
  Forte, and Manganelli]{Giudice17c}
Del~Giudice, V.; De~Paola, P.; Forte, F.; Manganelli, B.
\newblock {Real Estate appraisals with Bayesian approach and Markov Chain
  Hybrid Monte Carlo Method: An Application to a Central Urban Area of Naples}.
\newblock {\em Sustainability} {\bf 2017}, {\em 9},~2138.

\bibitem[White(1980)]{white}
\textls[-25]{White, H.~A Heteroskedasticity-Consistent Covariance Matrix Estimator and a
  Direct Test for Heteroskedasticity.
\newblock {\em Econometrica} {\bf 1980}, {\em 48},~817--838.}

\bibitem[MacKinnon and White(1985)]{MACKINNON}
MacKinnon, J.G.; White, H.
\newblock Some heteroskedasticity-consistent covariance matrix estimators with
  improved finite sample properties.
\newblock {\em J. Econom.} {\bf 1985}, {\em 29},~305--325.

\bibitem[Frank \em{et~al.}(2000)Frank, Trigg, Holmes, and Witten]{Frank00}
Frank, E.; Trigg, L.; Holmes, G.; Witten, I.H.
\newblock {Technical Note: Naive Bayes for Regression}.
\newblock {\em Mach. Learn.} {\bf 2000},~{\em 41},~5--25.

\bibitem[Smola and Sch\"{o}lkopf(2004)]{Smola04}
Smola, A.J.; Sch\"{o}lkopf, B.
\newblock {A Tutorial on Support Vector Regression}.
\newblock {\em Stat. Comput.} {\bf 2004}, {\em 14},~199--222.

\bibitem[Geurts \em{et~al.}(2006)Geurts, Ernst, and Wehenkel]{Geurts06}
Geurts, P.; Ernst, D.; Wehenkel, L.
\newblock {Extremely Randomized Trees}.
\newblock {\em Mach. Learn.} {\bf 2006}, {\em 63},~3--42.

\bibitem[Pedregosa \em{et~al.}(2011)Pedregosa, Varoquaux, Gramfort, Michel,
  Thirion, Grisel, Blondel, Prettenhofer, Weiss, Dubourg, Vanderplas, Passos,
  Cournapeau, Brucher, Perrot, and \'{E}douard Duchesnay]{Pedregosa11}
Pedregosa, F.; Varoquaux, G.; Gramfort, A.; Michel, V.; Thirion, B.; Grisel,
  O.; Blondel, M.; Prettenhofer, P.; Weiss, R.; Dubourg, V.; Vanderplas, J.;
  Passos, A.; Cournapeau, D.; Brucher, M.; Perrot, M.; \'{E}douard Duchesnay.
\newblock {Scikit-learn: Machine Learning in Python}.
\newblock {\em J. Mach. Learn. Res.} {\bf 2011}, {\em
  12},~2825--2830.

\bibitem[Liu \em{et~al.}(2017)Liu, Wang, Liu, Zeng, Liu, and Alsaadi]{Liu17}
Liu, W.; Wang, Z.; Liu, X.; Zeng, N.; Liu, Y.; Alsaadi, F.
\newblock {A Survey of Deep Neural Network Architectures and their
  Applications}.
\newblock {\em Neurocomputing} {\bf 2017}, {\em 243},~11--26.

\bibitem[Kleine-Deters \em{et~al.}(2017)Kleine-Deters, Zalakeviciute, Gonzalez,
  and Rybarczyk]{Deters17}
Kleine-Deters, J.; Zalakeviciute, R.; Gonzalez, M.; Rybarczyk, Y.
\newblock {Modeling PM2.5 Urban Pollution Using Machine Learning and Selected
  Meteorological Parameters}.
\newblock {\em J. Electr. Comput. Eng.} {\bf 2017}, {\em
  2017},~5106045.

\end{thebibliography}
\end{document}